\documentclass[fleqn,usenatbib]{mnras}

\usepackage[utf8]{inputenc} 
\usepackage[T1]{fontenc}
\usepackage{ae,aecompl}
\usepackage{tabularx}
\usepackage{lineno}
\usepackage{ulem}

\usepackage{epsf,palatino,url,wrapfig,array,setspace,amsmath,amssymb,fancyhdr,lscape,appendix,rotating}
\usepackage{graphicx}
\usepackage{longtable}
\usepackage{amssymb}
\usepackage{xcolor}
\usepackage{color}
\usepackage{lipsum}
\usepackage{textcomp}
\usepackage{subcaption}

\usepackage{pifont}
\usepackage{multirow}
\usepackage[para,online,flushleft]{threeparttable}

\newsavebox{\measurebox}

\usepackage{enumerate}
\usepackage{placeins}
\usepackage{float}
\usepackage{soul}
\usepackage{blindtext}

\usepackage{tikz,hyperref}
\usepackage{academicons}
\definecolor{lime}{HTML}{A6CE39}
\DeclareRobustCommand{\orcidicon}{%
	\begin{tikzpicture}
	\draw[lime, fill=lime] (0,0) 
	circle [radius=0.16] 
	node[white] {{\fontfamily{qag}\selectfont \tiny ID}};
	\draw[white, fill=white] (-0.0625,0.095) 
	circle [radius=0.007];
	\end{tikzpicture}
	\hspace{-2mm}
}

\foreach \x in {A, ..., Z}{%
	\expandafter\xdef\csname orcid\x\endcsname{\noexpand\href{https://orcid.org/\csname orcidauthor\x\endcsname}{\noexpand\orcidicon}}
}

\newcommand{\orcid}[1]{\href{https://orcid.org/#1}{\textcolor[HTML]{A6CE39}{\orcidicon}}}

\newcommand{\fermi}{{\it Fermi}\xspace}

\newcommand{\eqb}{\begin{eqnarray}}
\newcommand{\eqe}{\end{eqnarray}}

\newcommand{\eq}[2]{\begin{equation} \label{eq:#1} #2 \end{equation}}

\definecolor{frenchblue}{rgb}{0.0, 0.45, 0.73}
\definecolor{burgundy}{rgb}{0.5, 0.0, 0.13}
\definecolor{royalblue}{RGB}{65,105,225}
\definecolor{darkspringgreen}{rgb}{0.0, 0.65, 0.27}
\newcommand{\xmark}{\textcolor{burgundy}{\ding{55}}}%
\newcommand{\cmark}{\textcolor{darkspringgreen}{\ding{51}}}%
\newcommand{\qmark}{\textcolor{blue}{\emph{?}}}

\title[BlaVar]{A numerical study of long-term multi-wavelength blazar variability}

\author[M. Polkas et al.]
{M.~Polkas\orcid{0000-0002-8889-2167}$^{1}$\thanks{E-mail: markpolkas@phys.uoa.gr}, M.~Petropoulou\orcid{0000-0001-6640-0179}$^1$\thanks{E-mail:mpetropo@phys.uoa.gr},
G.~Vasilopoulos\orcid{0000-0003-3902-3915}$^{2,3}$,
A.~Mastichiadis\orcid{0000-0001-5217-4801}$^1$, \newauthor 
M.~C.~Urry\orcid{0000-0002-0745-9792}$^2$,
P.~Coppi$^2$, C.~Bailyn$^2$
\\
$^1$Department of Physics, National \& Kapodistrian University of Athens, Athens,Greece \\ 
$^2$Department of Astronomy, Yale University, PO Box 208101, New Haven, CT 06520-8101, USA \\
$^3$Université de Strasbourg, CNRS, Observatoire astronomique de Strasbourg, UMR 7550, 67000, Strasbourg, France
}

\date{Accepted XXX. Received YYY; in original form ZZZ}

\pubyear{2021}

\begin{document}

\pagerange{\pageref{firstpage}--\pageref{lastpage}}
\maketitle

\begin{abstract}
Decade-long monitoring of blazars at optical and infrared (OIR) wavelengths with the Small and Moderate Aperture Research Telescope System (SMARTS) in Chile and in $\gamma$-rays with the  \fermi \, Large Area Telescope (LAT) has enabled the systematic study of their multi-wavelength long-term variability. In this work we investigate, from a theoretical perspective, the long-term variability properties of blazar emission by introducing an observationally motivated time-dependence to four main parameters of the one-zone leptonic model: injection luminosity of relativistic electrons, strength of magnetic field, Doppler factor, and external photon field luminosity. For the first time, we use both the probability density function and the power spectral density of  the 10 year-long \fermi-LAT light curves to create variation patterns for the model parameters.  Using as test beds two bright blazars from the SMARTS sample (PKS 2155-304 and 3C~273), we compute 10 year-long OIR, X-ray, and $\gamma$-ray model light curves for different varying parameters. We compare the findings of our theoretical investigation with multi-wavelength observations using various measures of variability. While no single-varying parameter simulation can explain all multi-wavelength variability properties, changes in the electron luminosity and external radiation field in PKS~2155-304 and 3C~273, respectively, can account for most of them. Our results motivate future time-dependent studies with coupling between two or more physical parameters to describe the multi-wavelength long-term blazar variability.
\end{abstract}




\section{Introduction}\label{sec:intro}
Blazars are a subclass of active galactic nuclei (AGN) with relativistic jets closely aligned to our line of sight \citep{Urry1995} which are powered by accretion onto a central supermassive black hole \citep{Begelman1984}. They are the most powerful persistent astrophysical sources of non-thermal electromagnetic radiation in the Universe, with spectral energy distributions (SEDs) spanning $\sim15$ decades in energy, from radio frequencies up to high-energy $\gamma$-rays. 

Blazars are also characterized by flux variability that is frequency-dependent and manifests in a variety of timescales, ranging from minutes to years
\citep[for a review, see][]{Boettcher2019}. Short-duration $\gamma$-ray variability in particular has drawn a lot of attention, as it is one of the few blazar observables that can put to test and constrain theoretical models of $\gamma$-ray production in jets \citep[e.g.][]{MM08, Giannios2010, Ackermann2016, Petropoulou2017,Aharonian2017}. In addition to the very short variability timescales, a major challenge for our understanding of the broadband blazar emission is that flux-flux correlations between different energy bands do not show a consistent behaviour. In fact, even the same source can exhibit correlated and uncorrelated inter-band flux variability between different observation periods and/or on different timescales \citep[e.g.][]{Acciari}. 

From a theoretical perspective, inter-band flux correlations are naturally expected when the broadband emission originates from the same region in the jet and from the same particle population. The energy-dependent cooling (and/or escape) timescales of the radiating particles can also lead to time lags between different energy bands \citep[e.g.][]{Takahashi96,Chia99,Bottcher10,Hovatta2015}. For example, one-zone leptonic emission models of high-peaked BL Lac objects (HBL) like PKS~2155-304, predict strong correlation between X-rays and very high-energy $\gamma$-rays with small time lags, as both emissions are produced by electrons of similar energies \citep[][]{MK97,Maraschi08}. For flat spectrum radio quasars (FSRQs) like 3C 273, the same models postulate that the GeV $\gamma$-rays are  produced by inverse Compton scattering of ambient optical-near-infrared (OIR) photons by the synchrotron-emitting electrons in the jet \citep[][]{Dermer93}. Thus, GeV $\gamma$-rays and OIR wavelengths are expected to be correlated (or anti-correlated) with an intra-day time lag, while the strength of the correlation will depend on the relative contribution of the jet and the external photon fields (e.g. accretion disk, BLR, torus) to the OIR emission. In the context of one-zone leptonic models, different strengths of the correlation (and non-zero time lags) are expected, if temporal variations on more than one model parameters, such as magnetic field strength and Doppler factor, are considered \citep[e.g.][]{Kraw02,Katarzynski2005,Mast2013}. To extract physical information from the observed phenomenology on flux variability, we need to confront model predictions with long-term multi-wavelength variability patterns.

From an observational perspective, detailed studies of correlated flux variability in blazars are often hindered by the poor sampling of the light curves. In this regard, the Large Area Telescope (LAT) on board the \textit{Fermi} Gamma-Ray Space Telescope is a unique instrument due to its near-continuous $\gamma$-ray monitoring of blazars~\citep[][]{LAT}. With an $\sim11$ year-long operation period, \fermi-LAT produced a large sample of long-term blazar $\gamma$-ray light curves with regular sampling, which enabled cross-correlation studies of $\gamma$-ray light curves with radio and/or optical light curves \citep[e.g.][]{Wehrle2012,Rani2013, Hovatta2014,Williamson2016,Liodakis2019}. 
In this type of studies, the quality of light curves at lower wavelengths is also equally important. The meter-class telescopes of the Small and Moderate Aperture Research Telescope System (SMARTS) have produced good quality OIR light curves with regular cadence for a large sample of bright southern $\gamma$-ray blazars as part of the Yale/SMARTS blazar monitoring program\footnote{\url{http://www.astro.yale.edu/smarts/glast/home.php}}.

Most of the theoretical studies so far have been focusing on the modelling of flaring events of blazars by adopting one or two time-dependent parameters of the model  \citep[e.g.][]{MK97, Kraw02, Katarzynski06, Katarzynski07, Asano2015, glx2, Rajput20}. For example, \citet{MK97} produced flaring states by varying either the electron compactness (which is a  dimensionless measure of the electron injection luminosity) or the magnetic field with application to an observed flare in X-rays and TeV $\gamma$-rays of Mkn~421. \citet{Kraw02} studied the effects of a variable electron compactness and Doppler factor while trying to explain the  observed $\gamma$-ray variability of Mkn~501. \citet{Bottcher14}, using single pulses to perturb the electron luminosity, modeled the flaring of FSRQs, while considering the acceleration of electrons by the Fermi II process. More recently, \citet{glx2} have studied the long-term variability in the one-zone leptonic model using generic parameter values  of blazars and a red-noise power spectrum distribution for the variations of the electron luminosity, electron spectral index, and  magnetic field strength.

In this work we investigate the long-term variability properties of blazar emission by introducing an observationally motivated time-dependence to four main parameters of the one-zone leptonic model: electron compactness, strength of magnetic field, Doppler factor and external photon field compactness.
For the first time, we use both the probability density function (PDF) and the power spectral density (PSD) of  the observed multi-year  \fermi-LAT light curves, to create synthetic $\gamma$-ray light curves and variation patterns for the model parameters in order to  simulate the long-term multi-wavelength flux variability. Our goal is to understand the cause of the observed long-term variability properties at O/IR wavelengths, X-rays, and $\gamma$-rays. To do so, we compare the findings of our theoretical investigation with observations of two bright blazars from the SMARTS sample: the HBL PKS~2155-304 \citep[$z=0.116$,][]{pkshess}, which is known for its very fast flare at TeV energies \citep{Aharonian07} and its weak correlation between optical and GeV $\gamma$-rays at lower flux states \citep{AtomHess}, and the well-known FSRQ 3C~273 \citep[$z=0.158$,][]{Schmidt3c}, whose optical/ultraviolet (UV) spectrum shows a prominent excess of emission interpreted as a contribution from a luminous accretion disc \citep{Ulrich81, Soldi}. 

This paper is structured as follows. In Section~\ref{sec:code} we describe the one-zone leptonic model and the numerical approach used to compute time-dependent multi-wavelength photon spectra.
In Section~\ref{sec:blavar} we describe how we model long-term flux variability in the context of the one-zone leptonic model of blazar emission. In Section~\ref{sec:results} we apply our methodology to the FSRQ 3C 273 and the HBL PKS 2155-304 in order to study the long-term variability in different $\gamma$-ray emission scenarios. 
We then compare our simulation results against observations of both targets in OIR wavelengths and GeV $\gamma$-ray energies using various diagnostics. In Section~\ref{sec:discussion} we discuss the shortcomings of our model and the physical implications of our results. We conclude in Section~\ref{sec:conclusions} with a summary of our main findings.

\section{The one-zone leptonic model}\label{sec:model}
To compute the multi-wavelength blazar emission we adopt the one-zone framework as described in \citet{MK95}, according to which relativistic electrons and positrons (simply referred to as electrons) are injected into a spherical volume of constant radius $R$ (blob) that contains a tangled magnetic field of strength $B$. Particles subsequently cool via synchrotron and inverse Compton processes. Henceforth, unprimed quantities are measured in the rest frame of the blob, while we use the subscript ``obs'' to refer to quantities measured in the observer's frame. The blob is moving with a Lorentz factor $\Gamma$ with respect to an observer at angle $\theta_{\rm obs}$ \citep[see][for relativistic bulk motion in AGN]{Ghisellini93}. The radiation in the observer's frame is Doppler boosted, with the Doppler factor defined as $\delta = \Gamma^{-1} (1-\beta \cos(\theta_{\rm obs}))^{-1}$ and $\beta = \sqrt{1-1/\Gamma^2}$. 

A pre-accelerated population of electrons is injected into the blob with a power-law distribution with slope $p$ between a minimum and a maximum Lorentz factor, denoted respectively as $\gamma_{\min}$ and $\gamma_{\max}$. Electrons and photons are allowed to leave the region on  a characteristic timescale $t^{\rm e}_{\rm esc} = t^\gamma_{\rm esc}=t_{\rm cr}$, where $t_{\rm cr}=R/c$ is the light crossing time. In general, the injection luminosity is a time-dependent quantity and it is defined as
\eq{Le}{L_{\rm e}(t)=m_{\rm e} c^2\int_{\gamma_{\min}}^{\gamma_{\max}} {\rm d}\gamma\, Q^{\rm inj}_{\rm e}(\gamma, t)(\gamma-1),}
where $\gamma$ the Lorentz factor of the electrons and $Q^{\rm inj}_{\rm e}(\gamma,t)$ is the injection rate of relativistic electrons with Lorentz factor $\gamma$ at a specific time $t$. The normalization of the injection rate can be expressed in terms of the electron compactness,
\eq{le}{l_{\rm e}=\frac{3\,\sigma_{\rm T}\,L_{\rm e}}{4\pi\, m_{\rm e} \, c^3\, R},}
which is a dimensionless measure of the electron injection luminosity, $L_{\rm e}$. We also define the magnetic field compactness as $l_{\rm b} = \sigma_{\rm T} R u_{\rm B}/m_{\rm e} c^2$, where $u_{\rm B}=B^2/8\pi$ the magnetic field energy density in blob frame. 

Upon their injection to the blob, electrons can lose energy (cool) through various radiative processes. The main competing cooling processes for relativistic electrons in the jet are synchrotron radiation and inverse Compton scattering (ICS) on the produced synchrotron photons \citep[synchrotron self Compton; SSC][]{Jones74} and/or any external photons from, e.g. the accretion disk, BLR or torus \citep[external Compton scattering; ECS][]{Dermer92, Dermer93, Sikora94}. 

For ECS scenarios, we approximate the energy distribution of the ambient photon field with a grey body. For an isotropic BLR with energy density $u_{\rm ext, \rm obs}$ and effective black-body temperature $T_{\rm ext,\rm obs}$ (in the AGN frame), the two relevant model parameters are the co-moving temperature $T_{\rm ext}= \Gamma T_{\rm ext,\rm obs}$ and external photon compactness $l_{\rm ext} = \sigma_{\rm T} R \Gamma^2 u_{\rm ext,\rm obs}/m_{\rm e} c^2$.
Direct radiation from the accretion disk is not considered an important source of seed photons for ECS; this is a valid assumption as long as the blob lies at a large enough distance from the disk \citep[see][]{Dermer95,Ghisellini09}. Besides the accretion disk, other external radiation fields of the AGN, such as the infrared emission of the torus, are also assumed to have negligible contribution to the SED.

\subsection{Numerical approach}\label{sec:code}
For a given set of parameter (a list of which is presented in Table~\ref{tab:params}), we solve the kinetic equations describing the evolution of the differential electron ($n_{\rm e}$) and photon ($n_\gamma$) distributions inside the blob, using an updated version of the numerical code of \citet{MK95}. This uses the full expression for the Klein-Nishina regime at the ICS losses \citep{BG70} and accepts as an input time-varying parameters
\eq{kinel}{\frac{\partial n_{\rm e}(\gamma,t)}{\partial t}+\frac{n_{\rm e}(\gamma,t)}{t^{\rm e}_{\rm esc}}=Q_{\rm e}(n_{\rm e},n_\gamma,\gamma,t)+\mathcal{L}_{\rm e}(n_{\rm e},n_\gamma,\gamma,t)}
\eq{kinph}{\frac{\partial n_\gamma(x,t)}{\partial t}+\frac{n_\gamma(x,t)}{t^{\gamma}_{\rm esc}}=Q_\gamma(n_{\rm e},n_\gamma,x,t)+\mathcal{L}_\gamma(n_{\rm e},n_\gamma,x,t),} 
where $x$ is the photon energy in units of $m_{\rm e} c^2$. 

The mathematical operators $Q$ and $\mathcal{L}$ appearing in the equations above, denote respectively source and loss terms of electrons and photons. The electron loss term, $\mathcal{L}_{\rm e}(n_{\rm e},n_\gamma,\gamma,t)$, includes energy losses due to synchrotron radiation and inverse Compton scattering on synchrotron photons and on external photons, if applicable. The synchrotron emission is accompanied by synchrotron self absorption (SSA), which is included in the $\mathcal{L}_\gamma$ term; this process suppresses the production of low-energy photons (e.g.  at GHz frequencies and below). The $\gamma\gamma$ pair production, which acts as a loss term for photons and a source term for electrons, is considered in the terms $\mathcal{L}_{\gamma}$ and $Q_{\rm e}$. 
Electron-positron annihilation is also included in the operators $Q_{\gamma}$ and $\mathcal{L}_{\rm e}$, but is a negligible process for parameters relevant to blazar emission.
Finally, we assume that once the electrons cool down to $\gamma\simeq 1$, they escape the emitting region before they have time to accumulate, so no Compton down-scattering of photons  by cold electrons is taken into account.

Equations (\ref{eq:kinel}) and (\ref{eq:kinph}) can be solved to derive both steady-state and time-dependent solutions. For the former case, which is relevant for the description of the time-average blazar SED, all parameters are considered to be constant in time, including $L_{\rm e}$ (see equation~\ref{eq:Le}). In this case, the system reaches a steady state after several light crossing times, with the exact time needed depending on source parameters. If at least one parameter in equations (\ref{eq:kinel}) and (\ref{eq:kinph}) has explicit dependence on time, neither the photon nor the electron distribution reaches a constant density. In this case, the temporal profile of the escaping photon luminosity (i.e., light curve) is determined by the particle injection, cooling and escape rates. Our assumption that particles leave the emission region as quickly as photons ($t_{\rm esc}^{\rm e}=t_{\rm esc}^\gamma= R/c$) implies that changes in the photon luminosity will track closely changes in the injection rate, with any time lags between light curves depending on different cooling times~\citep[e.g.][]{Chia99, MM08, Mast2013}.

Although the modelling of flux variability in blazars can be approximated by a series of steady states under certain assumptions, a fully time-dependent calculation is more appropriate for our study. In the following section, we describe in detail our methods for modelling flux variability in blazars.

\begin{table}
\centering
\caption{Input parameters of the one-zone leptonic model for blazar emission.}
\begin{threeparttable}
\begin{tabular}{ll}
\hline
Symbol & Description\\
\hline\hline
$R$	&   Blob radius\\
$B$\tnote{\ddag}	& Blob magnetic field strength \\
\hline
$\delta$\tnote{\ddag}	& Doppler factor \\
$\Gamma$\tnote{\dag}	& Bulk Lorentz factor \\
$\theta_{\rm obs}$    & Observer's angle \\
\hline
$T_{\rm ext}$\tnote{\dag} & Temperature of external radiation field \\
$l_{\rm ext}$\tnote{\dag} \tnote{\ddag}& Compactness of external radiation field \\
\hline
$\gamma_{\min} $	& 	Minimum electron Lorentz factor  \\
$\gamma_{\max}$	& Maximum electron Lorentz factor \\
$p$	&  Power-law slope of injected electrons 	      \\
$l_{\rm e}$\tnote{\ddag}	&   Electron injection compactness 
\\
$t^{\rm e}_{\rm esc}\tnote{*}$ &   Electron escape time  \\
\hline
$z$  & Redshift \\
$D_{\rm L}$  &  Luminosity distance\\
\hline \hline
\end{tabular}
\begin{tablenotes}
\item[\dag] Parameters used only in ECS scenarios.\\
\item[\ddag] Time-varying parameters used for this study.\\ 
\item[*] This is fixed to the light crossing timescale, $R/c$, in all simulations.
\end{tablenotes}
\end{threeparttable}
\label{tab:params}
\end{table}

\section{modelling flux variability} \label{sec:blavar}

If all parameters are constant in time, the system reaches a steady state, except for cases involving leptonic radiative instabilities, such as the automatic photon quenching scenarios \citep[][]{StaKirk, Petropoulou11}. To model the variability of blazar emission we consider the simplest scenario where temporal variations are imposed on one physical parameter. Each time-varying parameter has a different impact on the simulated SEDs, as we will demonstrate in Section \ref{sec:results}. We choose to vary parameters that impact directly flux variability. By varying separately these four main parameters of the one-zone leptonic model, namely $l_{\rm e},\, B,\, \delta$ and $l_{\rm ext}$, we can study a variety of blazar multi-wavelength variability patterns, while being able to isolate and study in detail the effects of each physical parameter.  A qualitative discussion about physical processes giving rise to the assumed variability of the model parameters can be found in Section \ref{sec:discussion}. 

\subsection{The method}\label{sec:method}

\begin{figure}
\centering
\includegraphics[width=0.8\linewidth]{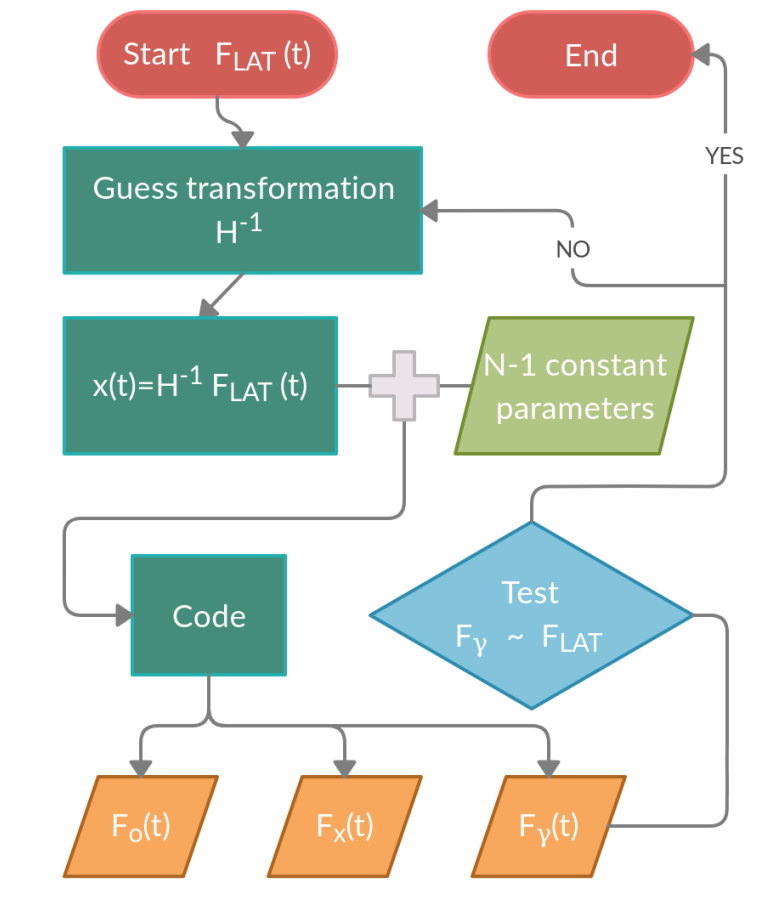}
\caption{Logical flow diagram showing the method used to induce flux variability in the one-zone model for blazar emission. We calculate a time series for a model parameter, $x(t)$, by applying a mathematical operator $H^{-1}$ (defined in equation~\ref{eq:normx}) to the synthetic LAT light curve, $\tilde{F}_{\rm LAT}$. Using $x(t)$ as an input to the code, we compute the time-dependent SED, and compare the produced $\gamma$-ray flux $F_\gamma(t)$ to the real $F_{\rm LAT}(t)$. If there are large differences, we modify the free parameters of the mathematical operator, and repeat the process. Otherwise, we use the output of the code to compute the various diagnostics introduced in Section~\ref{sec:analysis}}.
 \label{fig:inoutMK}
\end{figure}

In earlier works on blazar variability, the time series of one or more model parameters were typically computed using analytical expressions \citep[e.g. Lorentzian or rectangular pulses,][]{MK97,Dermer93,Sikora01} or were generated from a pre-selected PSD describing for instance a red-noise process \citep[e.g.][]{Mast2013,glx2}. In this study, we use an alternative method for modelling variability, which is motivated by the long-term $\gamma$-ray observations of blazars with \fermi-LAT. 
For parameter values that lead to fast electron cooling or fast electron escape time scales, one can directly map the observed flux variability to temporal changes of certain model parameters. In other words, we can generate input time series for each model parameter  
such that the time-dependent output of the code  in $\gamma$-rays~($0.1-300$~GeV) mimics the observed light curve in that band.
These input time-series, when fed into the one-zone leptonic model, produce the time-dependent SEDs, which are analyzed in Section \ref{sec:results}.
Our methodology is described in detail in the following paragraphs, while the key steps are summarized in the form of a flowchart in Figure~\ref{fig:inoutMK}.

\subsection{Real and synthetic $\gamma$-ray light curves}
We use the real \fermi-LAT light curves, which consist of $\gamma$-ray flux measurements in the $0.1-300$ GeV energy range, $F_{\rm LAT}$, binned every one day, for a time interval of $T_{\rm obs}>3500$ days (MJD 54683 -  58329 for PKS~2155-304 and MJD 54683 -  58332  for 3C~273).
The daily binned \fermi-LAT light curves of PKS~2155-304 and 3C~273 are, respectively, from \citet{Yoshida2021} and \citet{Meyer19}. We treat as upper limits bins with a test statistic\footnote{The test statistic (TS) is defined as the difference in the maximum likelihood of a model with and without the source \citep{1996ApJ...461..396M}.} (TS) less than 4, which corresponds to a $2\sigma$ excess, and bins with 68\% uncertainties in flux larger than the flux itself. These selection criteria yield real $\gamma$-ray light curves with 1545 and 2018 data points for PKS~2155-304 and 3C~273, respectively.

\begin{figure}
\centering
\includegraphics[width=1.0\linewidth]{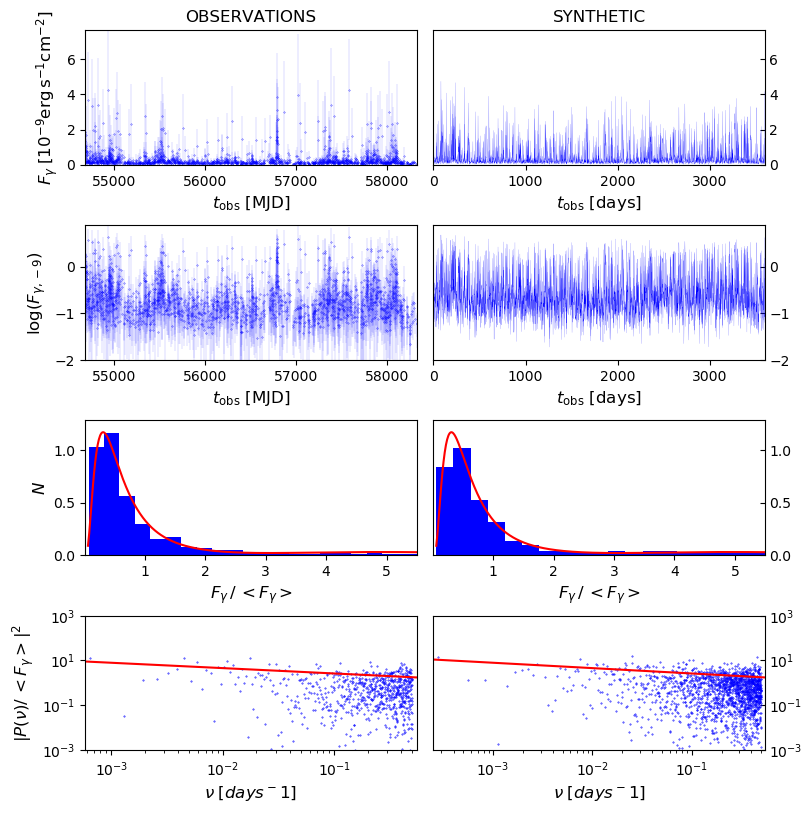}
\caption{Synthetic $\gamma$-ray light curve of PKS~2155-304 (top right panel) computed using the \citet{emmanou} method on the daily binned \fermi-LAT light curve with error bars indicating the $1\sigma$ uncertainties (top left panel). Panels in the second row from top show the $\gamma$-ray light curves in logarithmic scale.
The PDFs (third row from top) and PSDs (bottom row) are fitted with simple models (red lines) against the real and synthetic data (blue histogram and points). Each PDF is fitted with two log-normal distributions,
while each PSD is modelled as a broken power law \citep[see e.g.][]{epitropakis}.
}
\label{fig:DELCpks}
\end{figure}

\begin{figure}
\centering
\includegraphics[width=1.0\linewidth]{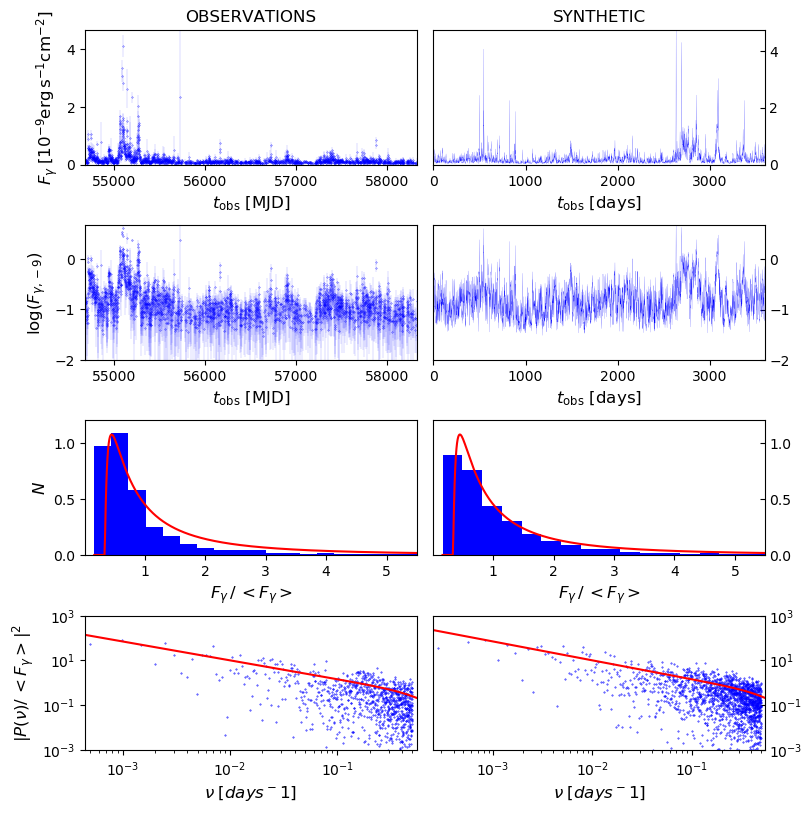}
\caption{Same as in Figure~\ref{fig:DELCpks} but for 3C~273. Here, the PDFs are fitted with one skewed log-normal distribution.}
 \label{fig:DELC3c}
\end{figure}
We then create for each object a synthetic light curve, $\tilde{F}_{\rm LAT}$, with the same timing properties as the observed \fermi-LAT light curve from the previous step (excluding the upper limits). For this purpose, we adopt the method of \citet{emmanou} that uses both the PDF and the PSD of the real LAT light curve to create synthetic $\gamma$-ray light curves. This method takes as an input a long continuous light curve. However, the observed \fermi-LAT light curve can have some gaps. To account for those, we can either ``glue'' together the continuous segments of the time series by omitting the gaps or we can interpolate between the gaps. We have checked that the former method does not introduce any stiff jumps to the synthetic light curves, and is preferred over the interpolation methods that are known to contaminate the power spectrum of a time series, as demonstrated by \citet{Wise99}. The new continuous synthetic time series has the same statistical properties with the observed \fermi-LAT light curve, while having a larger number of equally spaced in time data points (see Figures~\ref{fig:DELCpks} and \ref{fig:DELC3c}).

\subsection{Initial conditions}\label{sec:ss}
To set the initial conditions for the time-dependent calculations that follow, we compute a steady-state model for the blazar SED.
Using publicly available data from the \href{https://tools.ssdc.asi.it/}{Space Science Data Center (SSDC)}, we build time-filtered SEDs, each one comprised of observations taken within 2 year-long time intervals and spanning a 10 year-long period (2008-2018), similar to the duration of the real \fermi-LAT light curves used in our analysis. The purpose of using the 2 year-long data sets is to illustrate the variability of the source on those timescales, as the main goal of our simulations is to explain as much as possible long-term flux variations at different wavelengths. 

We should note that the time-filtered SEDs are not \textit{per se} time-averaged on a 2-year period. The $\gamma$-ray flux points in the $0.1-300$ GeV energy range, which are taken from the Second \fermi-LAT catalog \citep[2FGL,][]{2FGL} and Third \fermi-LAT catalog \citep[3FGL,][]{3FGL}, are indeed averaged over a 2-year and 4-year period, respectively. Meanwhile, X-ray and optical/UV observations usually capture a snapshot of the source within the 2-year time window. Simply taking the mean flux of these short observations would not provide a representative long-term average flux state at lower energies.

Therefore our primary goal is to find a steady-state model that describes the truly long-term average $\gamma$-ray spectrum well, while passing through the points of at least one of the 2-year data sets at lower energies. For 3C~273,  in particular, we use an additional constraint in the search of the steady-state model: its average jet contribution to the $\sim 10^{14.5}$~Hz flux should not exceed $\sim25\%$ of the average disk flux at the same frequency \citep{2020ApJ...897...18L}.

Although the selected steady-state blazar model is not the best-fit model to the data or a model for the average SED over 10 years, it is sufficient for setting the initial conditions for our calculations. We also discuss the impact of the selected steady-state model on our conclusions about broadband variability in Section~\ref{sec:discussion}.

\subsection{Parameter time series}
When the Doppler factor is the time-dependent parameter under consideration, equations (\ref{eq:kinel}) and (\ref{eq:kinph}) yield a steady-state solution in the blob comoving frame. This applies also to the ECS scenario, under the hypothesis that $\Gamma$, and thereby $l_{\rm ext}$, remain constant. Under these assumptions, variations in the Doppler factor are attributed to changes in the observer's angle. Because no changes occur in the blob frame, there is no need to
make Lorentz transformation of time between the observer's frame and the rest frame of the blob.
We therefore construct the time series $\delta(t_{\rm obs})$ from the synthetic $\gamma$-ray light curve $\tilde{F}_{\rm LAT}(t_{\rm obs})$
and use it to map fluxes from the comoving frame to the observer's frame (i.e., $F_{\nu_{\rm obs}}(t_{\rm obs})\propto \delta(t_{\rm obs})^{3+\alpha}F_{\nu}(t)$, where $\alpha$ is the spectral index and $F_{\nu}$ is the differential in frequency flux).
 
To model the time-dependence of $l_{\rm e}$, $B$ or $l_{\rm ext}$ we have first to transform $\tilde{F}_{\rm LAT}(t_{\rm obs})$ into the comoving frame of reference using the Doppler factor value of the steady-state blazar model, i.e., $t=\delta_0 t_{\rm obs}/(1+z)$, where $z$ is the redshift of the source.
The time series of parameter $x$ is constructed using a transformation of the form  $x = \hat{H}^{-1}\tilde{F}_{\rm LAT}$, where the operator $\hat{H}^{-1}$ is determined by the underlying physics in the blob and the $\gamma$-ray emission process (SSC or ECS). A power-law transformation
\eq{normx}{x(t_{\rm j})=x_0\left(\frac{\tilde{F}_{\rm LAT}(t_{\rm j})}{\langle \tilde{F}_{\rm LAT}(t_{\rm j})\rangle}\right)^{1/\sigma_{\gamma}} \cdot}
is a physically motivated choice \citep[][]{Kardashev,Dermer95,Zacharias10,PPM15}.
Here, $x_0$ is the steady-state value of the time-dependent parameter, $\tilde{F}_{\rm LAT}(t_{\rm j})$ and $\langle \tilde{F}_{\rm LAT}(t_{\rm j})\rangle$ are the
\fermi-LAT synthetic light curve and its time-average value respectively (with time measured in the comoving frame of the blob). 
The power-law index $\sigma_\gamma$ is assumed to be an integer number that is independent of the source's flux state (i.e. constant at all times). The adopted values for $\sigma_\gamma$ depend on the emission scenario (SSC or ECS) for the blazar under study and on the time-varying parameter (see Table~\ref{tab:sigmag}). An analytical approach 
provides a first good guess of the $\sigma_\gamma$ value to be used in equation~(\ref{eq:normx}) (see Appendix~\ref{sec:app} for more details). For instance, the $\gamma$-ray flux (in the LAT range) depends almost linearly on the electron luminosity (i.e.,  $\sigma_\gamma\simeq 1$) in the ECS scenario (assuming that SSC rarely dominates the ECS emission). On the contrary, the $\gamma$-ray flux in the SSC scenario scales roughly quadratically with the electron luminosity (i.e., $\sigma_{\gamma} \simeq 2$). The estimation of $\sigma_\gamma$ is more complicated for other time-varying parameters, like the magnetic field, as they affect the relative contribution of the synchrotron, SSC, and ECS emission processes to the spectrum, and cause spectral changes that must be taken into account. In fact, we will show later in Section~\ref{sec:results} that the assumption of a time-independent $\sigma_\gamma$ in some of our simulations should be reevaluated.

\begin{table}
    \caption{Time-varying parameters and values of the power-law index $\sigma_\gamma$ (see equation~\ref{eq:normx}) used for different blazar subclasses (see also Appendix~\ref{sec:app}).} 
    \centering
    \begin{threeparttable}
    \begin{tabular}{c c c}
    Parameter     &   \multicolumn{2}{c}{$\sigma_\gamma$}\\
    \hline
    & HBL & FSRQ \\
    \hline
    $l_{\rm e}$ & 2 & 1 \\
    $B$ & 1 & -1\\
    $\delta$ & 4 & 5\\
    $l_{\rm ext}$ & n/a & 1\\
    \hline
    \end{tabular}
    \end{threeparttable}
    \label{tab:sigmag}
\end{table}

\begin{figure}
\centering
\includegraphics[width=1.0\linewidth]{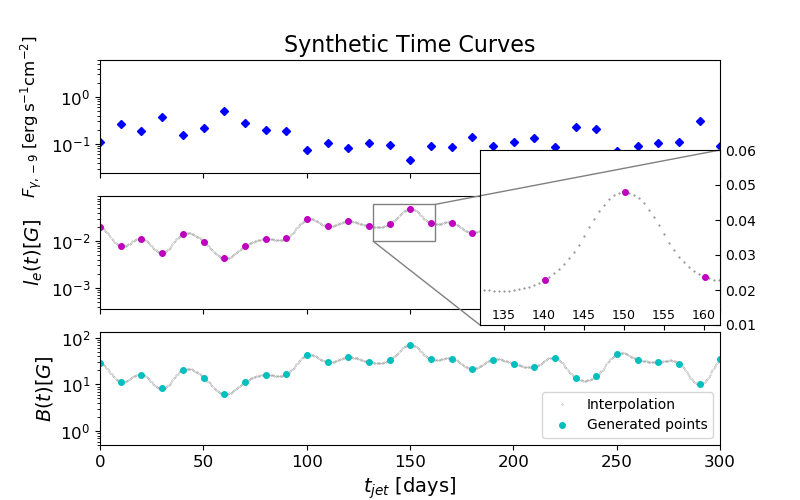}
\caption{Zoom in to a part of the full synthetic $\gamma$-ray light curve of 3C~273 (top panel). Middle and bottoms panels show the generated time series for $l_{\rm e}$ and $B$, respectively. Solid lines show the result of quadratic interpolation between generated points (in logarithmic space). Time is measured in the jet comoving frame.}
 \label{fig:timeseries}
\end{figure}

\subsection{Code description and outputs}\label{sec:analysis}
The time series used as an input to the code described in Section~\ref{sec:code} should be a smooth function of time, avoiding unreasonably steep changes in the second derivatives of that function. For this reason, we perform quadratic interpolation of $x(t)$ in logarithmic space at points that are equally spaced in time with $\Delta t=0.25\;t_{\rm cr} \simeq 0.95~{\rm d}\, (R/10^{16}~{\rm cm})$. This way we avoid oversampling, while having a temporal resolution  shorter than the quasi-stationary timescale that is $\sim 1\;t_{\rm cr}$. Figure~\ref{fig:timeseries} shows indicative examples of the interpolated time series for parameters used in the simulations of 3C~273. For the selected parameter values, the interval of flux points in the synthetic light curve is $\sim10$~days, while the number of interpolating points in between is $\sim16$. 

We numerically solve equations (\ref{eq:kinel}) and (\ref{eq:kinph}) using the interpolated time series $x(t)$, while keeping all other parameters fixed to their values obtained for the steady-state blazar model (see Table~\ref{tab:steady}). We finally derive the differential in energy photon and electron number densities in the comoving frame with a temporal resolution of $1\, t_{\rm cr}$. After transforming the photon spectra to the observer's frame, we compute light curves in X-rays by integrating over the $2-10$ keV energy range, in $\gamma$-rays by integrating in the $0.1-300$~GeV energy range of \fermi-LAT, and at OIR wavelengths. For the latter case,  we integrate the time-dependent photon spectra over three narrow  ranges roughly centered at the effective wavelengths of the $K$, $J$ and $B$ SMARTS filters (i.e. $1.7-4.5\,\mu$m, $0.7 - 1.7\,\mu$m, $0.25-0.7\,\mu$m). For FSRQ sources, we add to the non-thermal OIR emission from the jet the flux of the disk and BLR components.
The BLR, which is the primary contributor of external photons in the blob, is considered to have variable luminosity, to match the changes of external photon compactness $l_{\rm ext}$, when those are induced. For simplicity, we consider the disk component constant in all simulations.
Possible time lags between changes occurring in the BLR region and in the external radiation received by the blob are not taken into account to avoid introducing more free parameters to the problem.

Fluxes of the simulated OIR, X-ray and $\gamma$-ray light curves are averaged over 1-day bins. Looking at intra-day timescales would not yield meaningful information, since  the simulated fluxes are generated from time series with temporal resolution of 1 day. We then perform timing analysis on the daily-binned model light curves. More specifically,
\begin{itemize}
    \item we compute the coefficient of variation (CV) -- the ratio of the standard deviation to the mean -- of simulated light curves as a function of observing frequency.
    \item we compute the discrete correlation function \citep[DCF,][]{DCF} for three pairs of light curves (i.e., $J$-band versus $2-10$~keV X-rays, $J$-band versus $0.1-300$~GeV $\gamma$-rays, and $2-10$~keV versus $0.1-300$~GeV $\gamma$-rays).
    \item we create colour-brightness diagrams. For this purpose, we derive the $B-J$ colour and $J$ magnitude from our simulated light curves. The SMARTS $J$ filter is a near-infrared filter that is closest to optical wavelengths, so it fairly resembles the behaviour of the whole OIR range.
\end{itemize}

In addition to the timing analysis, we cross check the accuracy of the transformation $\hat{H}^{-1}$ (see equation~\ref{eq:normx}) by comparing the flux distributions  of the real \fermi-LAT light curves and simulated $\gamma$-ray light curves in the $0.1-300$~GeV energy range (see flowchart in Figure \ref{fig:inoutMK}).
Our results for the BL Lac PKS~2155-304  and the FSRQ 3C~273 are presented in Section~\ref{sec:results}.

%
%
%
%
%
%
%
%
%
%
\section{Results}\label{sec:results}
We perform three simulations of flux variability for PKS~2155-304 and four simulations for 3C~273 (see Table~\ref{tab:sigmag}). The steady-state model parameters, acquired as described in Section~\ref{sec:ss}, 
are provided in Table~\ref{tab:steady}. The selected values of $R$ and $\delta$ for the steady-state model of PKS~2155-304 are also in agreement with the shortest variability timescale detected in X-rays with  \textit{Suzaku} \citep{Zhang_2021}. To illustrate the effects of the baseline model (which is used as an initial condition for the time-dependent simulations) we derive two steady-state models for 3C~273: one that is synchrotron-dominated for $B_0=22$~G and another one that is ECS-dominated for $B_0=10$~G. In both cases, the contribution of the SSC component to the $\gamma$-ray flux of 3C~273 can be ignored for the biggest part of our simulations. Only in some extreme flaring states and for certain time-dependent parameters, the SSC flux can exceed that of the ECS component. 

In the simulations of 3C~273 we consider the BLR as the external photon field. Combining the BLR covering factor $f_{\rm BLR}=0.1$ \citep{Rees89} with the disk luminosity $L_{\rm disk}\simeq 5\times10^{46}$~erg s$^{-1}$ \citep{Malkan}, we obtain an estimate of the bolometric luminosity of the external photon field $L_{\rm ext,\rm obs}=2\times 10^{45} \;{\rm erg\,s^{-1}}$. The selected BLR radius $R_{\rm ext,\rm obs}\simeq 7\times 10^{17}$~cm is estimated using the scaling relation $R_{\rm ext,\rm obs}\propto L_{\rm disk}^{1/2}$ \citep{tavecchio08}. We adopt an effective black-body temperature of $T_{\rm ext, obs}\simeq 9\times10^3$~K which corresponds to an average photon energy of 2.3~eV in the AGN frame. Another source of external thermal radiation  could be that of the AGN torus with temperatures $\sim 100-1000$~K \citep[e.g.][]{cleary07}. In what follows, we will not include in our calculations the infrared radiation from the torus, since for typical radii  \citep[e.g.][]{honig07, Kishimoto11, Sobrino20} its energy density will be much smaller than that of the BLR. We will also neglect the direct irradiation from the accretion disc \citep{Dermer02}. This is a safe assumption as long as the emission region lies at distances $\gtrsim  0.01~{\rm pc} \,  (f_{\rm BLR}/0.1)^{-1/3} (L_{\rm disk}/10^{45}~{\rm erg s}^{-1})^{1/3} (M_{\rm BH}/10^9 M_{\sun})^{1/3}$ \citep{Dermer02, Sikora09}, where $M_{\rm BH}\sim (0.9-2.4)\times10^9 M_{\sun}$ is the black-hole mass of 3C~273 \citep{Peterson04, Paltani05}.

\begin{table}
\caption{Parameter values of the steady-state models for PKS~2155-304 and 3C~273.}
\centering
\begin{threeparttable}
\begin{tabular}{l c c}
\hline \hline
Parameter & PKS~2155-304 & 3C~273 \\ \hline \hline
$R\; ({\rm cm}) $ & $5\times10^{16}$ &  $	6.3\times10^{15}$\\
$B\;({\rm G}) $ & $	0.05 $  & $	22\, (10)$\tnote{\dag} \\
$\gamma_{\min} $ & $4\times10^3$  & $1$\\
$\gamma_{\max} $ & $5\times10^5$ & $3\times10^3$\\
$p $ & $	3.0 $ &  $	2.15 $ \\
$l_{\rm e}$ &  $1.8\times10^{-5}$  & $1.2\times10^{-2}$ \\
$l_{\rm ext}$ &   $-$   &	$3.3\times10^{-2}$     \\
$T_{\rm ext}\; ({\rm K})$ & $-$    & $ 3.1 \times 10^5$  \\
$\delta $ & $	30 $ & $	10 $ \\
$\Gamma $ & $	30 $ &  $	35 $  \\
$ \theta_{\rm obs} \; ({\rm deg})$  & $	1.9 $  & $ 4.0$\\
\hline
\end{tabular}
\begin{tablenotes}
Note -- $T_{\rm ext}, l_{\rm ext}$ are given in the comoving frame of the blob. \\
\item[\dag]The value in parenthesis, when all other parameters remain unchanged, defines an alternative steady state that we model separately.
\end{tablenotes}
\end{threeparttable}
\label{tab:steady}
\end{table}

The energy distribution of the external photon field is Doppler boosted to the blob comoving frame and then added to the energy distribution of non-thermal photons. The numerical code solves equations~(\ref{eq:kinel}) and (\ref{eq:kinph}) for the combined photon distribution, $n_{\gamma}$. In the $l_{\rm ext}$-varying simulations of 3C~273, the number density of external photons frequently dominates the non-thermal one in the soft X-rays (i.e. $\sim0.1-2$ keV in the observer’s frame). As a result, when we subtract the external component from the total photon number distribution, $n_\gamma$, to compute the non-thermal emission, we have to interpolate the latter typically over a decade in energy. This is indicated by a dashed-hatched region in the SED plots of 3C~273 that follow. Because the soft X-ray flux will depend on the interpolation, we do not use it for the DCF and CV analysis.

\begin{figure}
\centering
\includegraphics[width = 0.47 \textwidth]{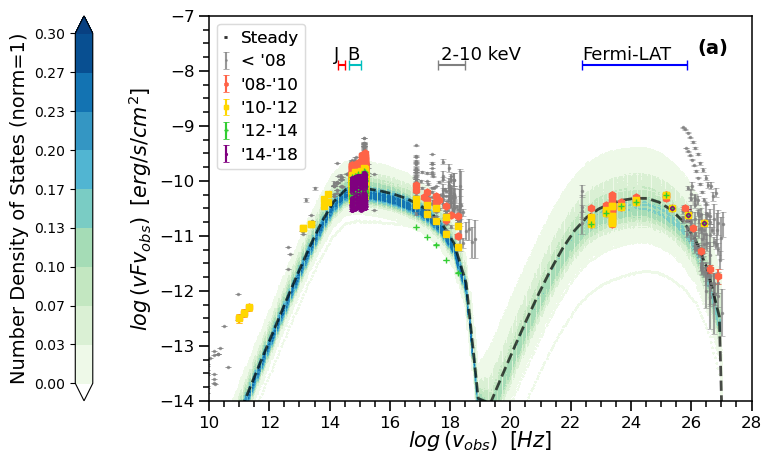}
\includegraphics[width = 0.47 \textwidth]{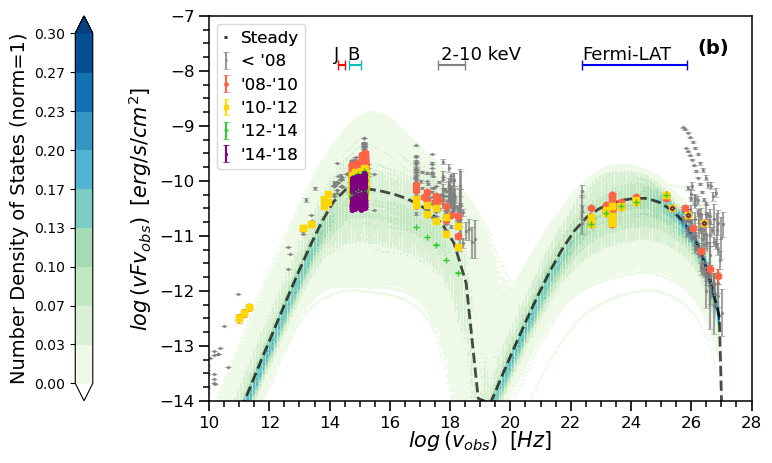}
\includegraphics[width = 0.47 \textwidth]{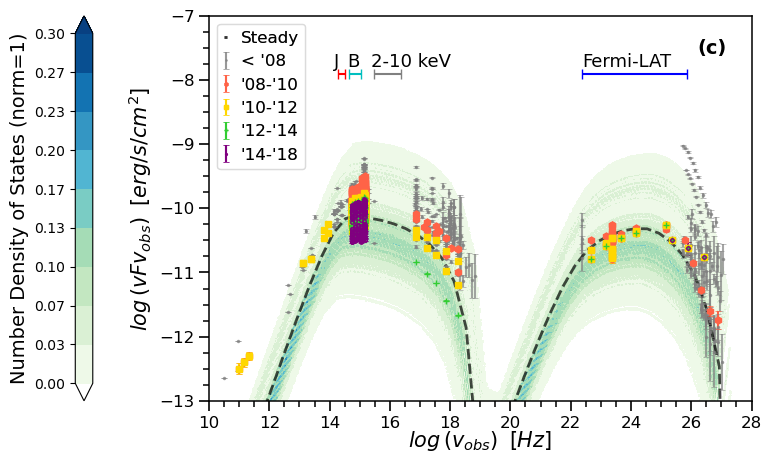}
\caption{Density map of flux states at different frequencies obtained from our long-term simulations of PKS~2155-304 with variable $l_{\rm e}$ (panel a), $B$ (panel b), and $\delta$ (panel c). The colour indicates the density of flux states normalized to unity (see colour bar). Darker colours  suggest that the source spends a larger fraction of simulation time in a particular flux state. The observed SED is compiled using publicly available data (symbols) from the the \href{https://tools.ssdc.asi.it/}{Space Science Data Center}. Observations prior to 2008 that do not coincide with the 10-year period of our timing analysis are shown in grey for comparison. The steady-state model is overplotted (dashed black line). 
Animations of the simulations can be found at  
\href{https://www.youtube.com/playlist?list=PL7Pc6hSbFNHcNJqVUc6_W_ds_6t49QTEI}{YouTube}.
}
\label{fig:vFvs_pks}
\end{figure}

\begin{figure*}
\centering
\includegraphics[width = 0.47 \textwidth]{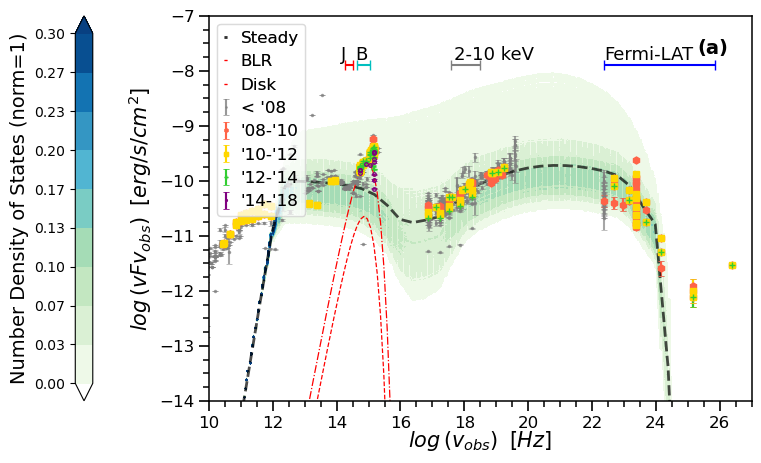}
\includegraphics[width = 0.47 \textwidth]{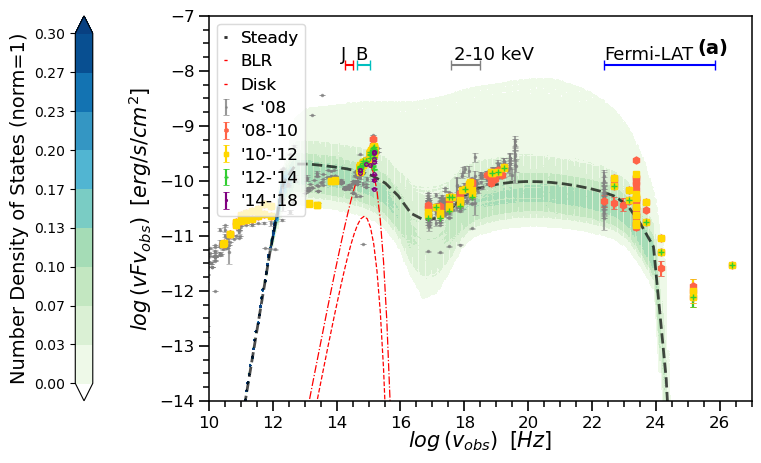}\\
\includegraphics[width = 0.47 \textwidth]{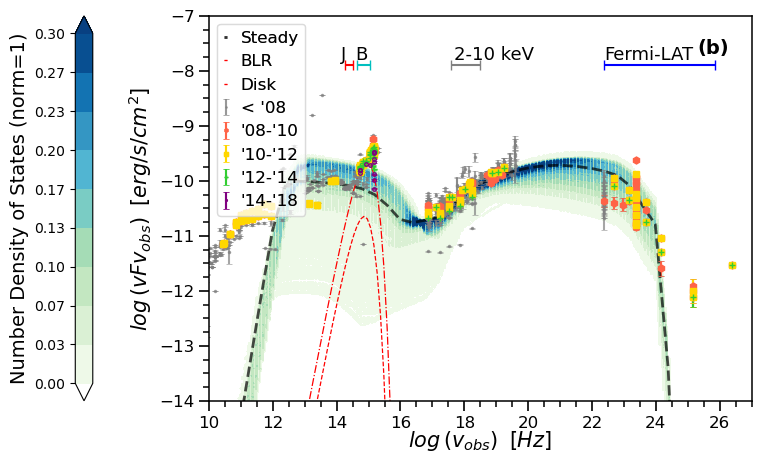}
\includegraphics[width = 0.47 \textwidth]{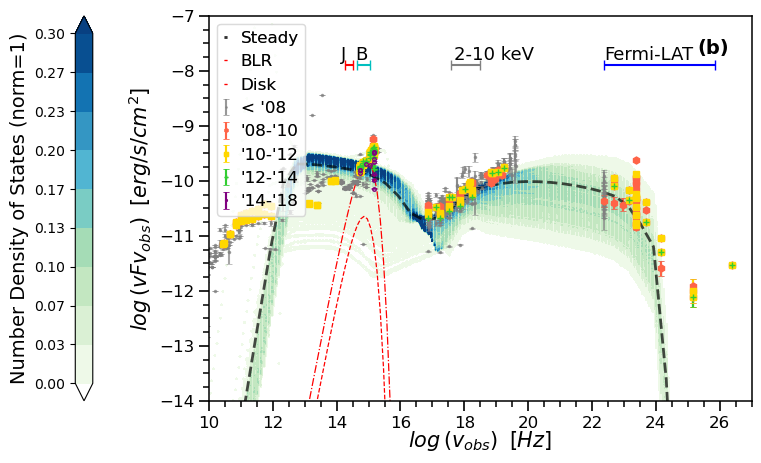}\\
\includegraphics[width = 0.47 \textwidth]{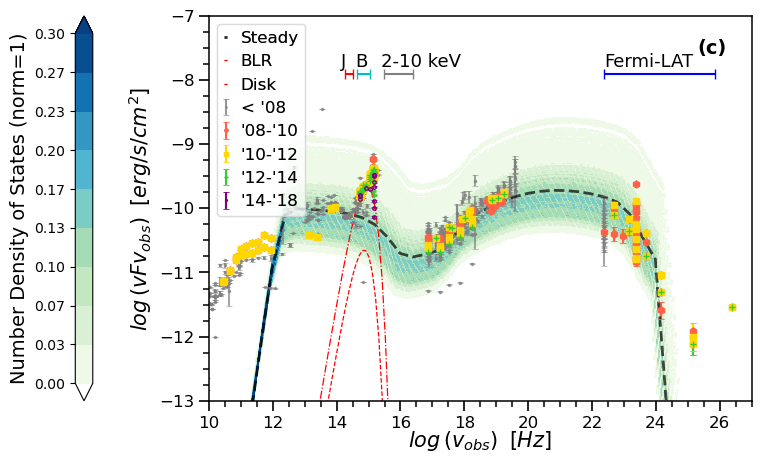}
\includegraphics[width = 0.47 \textwidth]{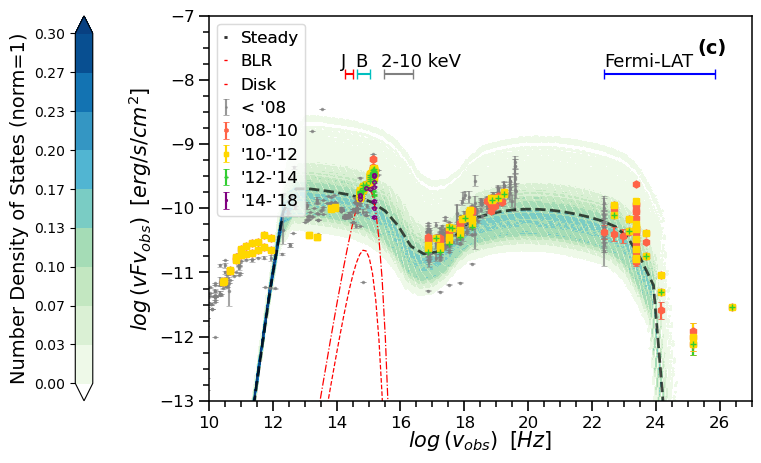}\\
\includegraphics[width = 0.47 \textwidth]{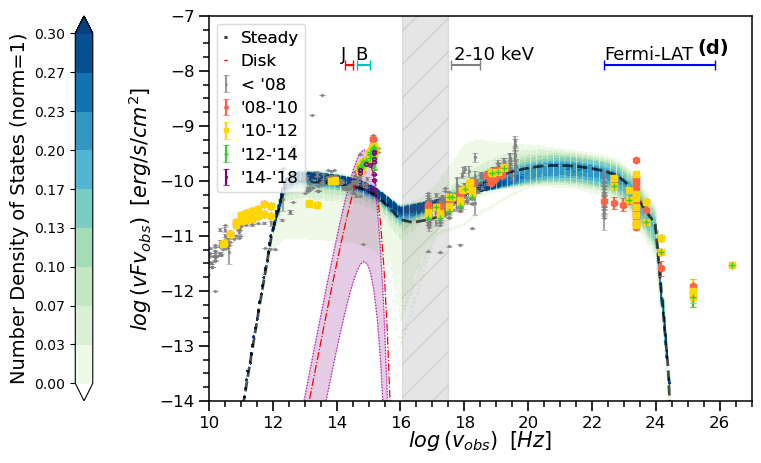}
\includegraphics[width = 0.47 \textwidth]{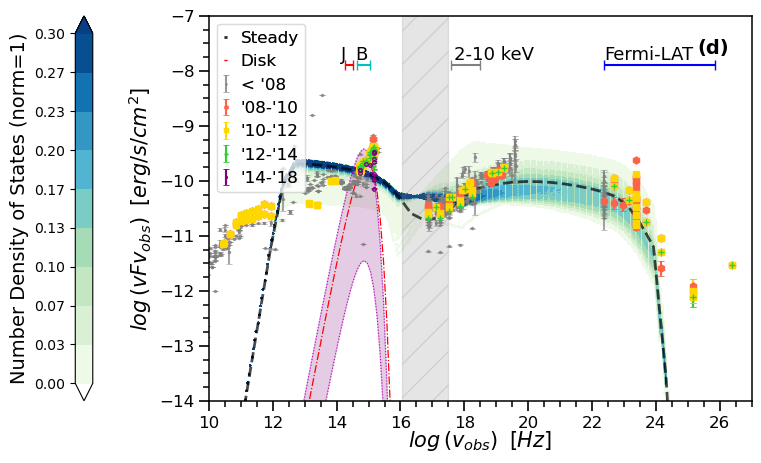}
\caption{Same as in Figure~\ref{fig:vFvs_pks} but for 3C~273. Results from long-term simulations with variable $l_{\rm e}$ (panel a), $B$ (panel b),  $\delta$ (panel c) and $l_{\rm ext}$ (panel d) are computed for two steady-state models with $B_0=10\;{\rm G}$ (left column) and $B_0=22\;{\rm G}$ (right column). In panels (d) the purple coloured region indicates changes in the BLR spectra, while the dashed-shaded region indicates the energy range where interpolation of the simulated spectra is necessary (see text in Section~\ref{sec:results} for more details). For completeness, we also show the accretion disk component that is modeled as a simple black body. 
Animations of the simulations can be found at \href{https://www.youtube.com/playlist?list=PL7Pc6hSbFNHcNJqVUc6_W_ds_6t49QTEI}{YouTube}.
}
\label{fig:vFvs_3c}
\end{figure*}
\subsection{Time-dependent SEDs}
Results from our time-dependent simulations for PKS~2155-304 and 3C~273 are presented in Figures \ref{fig:vFvs_pks} and \ref{fig:vFvs_3c}, respectively. The contour plots displayed in these figures indicate the probability of finding the blazar at a certain flux level at different  energies\footnote{The number of states for each blazar is not normalized separately for discrete energy bins, but for the whole energy range of the SED.} over the course of $\sim 10$~years, and serve as a first visual diagnostic of flux variability. Darker colours  indicate a high concentration of states, and as such, they are clustered around the time-average SED, which is similar to the steady-state SED model displayed with a dashed black line. In a particular energy band, the lack of dark colours  means that the flux in these energy ranges varies greatly within the simulation compared to other energy ranges that appear darker in colour (see e.g. the \fermi-LAT energy range in panels a-c of Figure~\ref{fig:vFvs_3c}). 

In PKS~2155-304 (Figure~\ref{fig:vFvs_pks}), the magnetic field time dependence (panel b) yields significant changes in the spectral shape of the low-energy hump of the SED. This is a direct consequence of the varying efficiency of the dominant electron cooling mechanism. The synchrotron cooling break frequency is sensitive to changes of the magnetic field ($\propto {B}^{-3}$), and it can therefore change by more than three orders of magnitude over the course of the simulation.  A larger part of the electron population is fast cooling for stronger magnetic fields, thus resulting in steeper synchrotron spectra. While extreme changes of the peak synchrotron frequency have been detected in certain BL Lac objects during flares, the peak is usually found to move to higher frequencies with increasing flux \citep[for Mkn~501, see e.g.][]{Pian98,Tavecchio01}, which is opposite to the model prediction. The  spectral shape in the \fermi-LAT energy band  is less strongly affected by changes in the magnetic field, since the SSC emission in any particular energy range has contributions from different parts of the electron and synchrotron photon energy distributions.
In the $l_{\rm e}$-varying and $\delta$-varying simulations, the electron cooling rate is constant (unless SSC becomes the dominant cooling mechanism). Thus, the observed flux/spectral changes are caused solely by changes in the luminosity or characteristic emitting frequency. Between the $l_{\rm e}$-varying and $\delta$-varying simulations, the latter produces stronger flux variability at all wavelengths (this will be demonstrated in a quantitative way in later in this section, see Figure~\ref{fig:allCVs}).

In 3C~273 (Figure \ref{fig:vFvs_3c}) the assumed time-dependence on $B$ (panel b) and $l_{\rm ext}$  (panel d)  affects the efficiency of the synchrotron and ECS cooling mechanisms. Depending on which process dominates the electron cooling, changes in one of the aforementioned model parameters may have an impact on the spectral shape of the synchrotron and Compton components of the SED. Electrons are cooling mostly due to synchrotron radiation in the $B_0=22$~G simulation, while inverse Compton scattering on BLR photons governs electron cooling in the $B_0=10$~G simulation most of the times. These differences are reflected on the broadband flux variability and the intensity of spectral changes, as it can be seen by comparing panels b (or panels d) on the left-hand and right-hand sides of the figure. In the ECS model, any changes in the cooling break energy of electrons can be imprinted on both the synchrotron and ECS components of the SED, unlike the SSC scenario where these are more subtle in the SSC spectrum. Hence, changes in the magnetic field can induce spectral changes in the \fermi-LAT energy range (compare also panels b in Figures \ref{fig:vFvs_pks} and \ref{fig:vFvs_3c}). In the ECS scenario for 3C~273, both $l_{\rm e}$ and $\delta$ have similar effects on flux variability as in SSC models (compare e.g. panels a in Figures \ref{fig:vFvs_pks} and \ref{fig:vFvs_3c}). The additional feature with respect to SSC models is the competition between the SSC and ECS processes, which contributes to the spectral variability in the X-ray and soft $\gamma$-ray bands in the case of the $l_{\rm e}$-varying simulations.

\begin{figure*}
\centering
\includegraphics[width = 1.0 \textwidth]{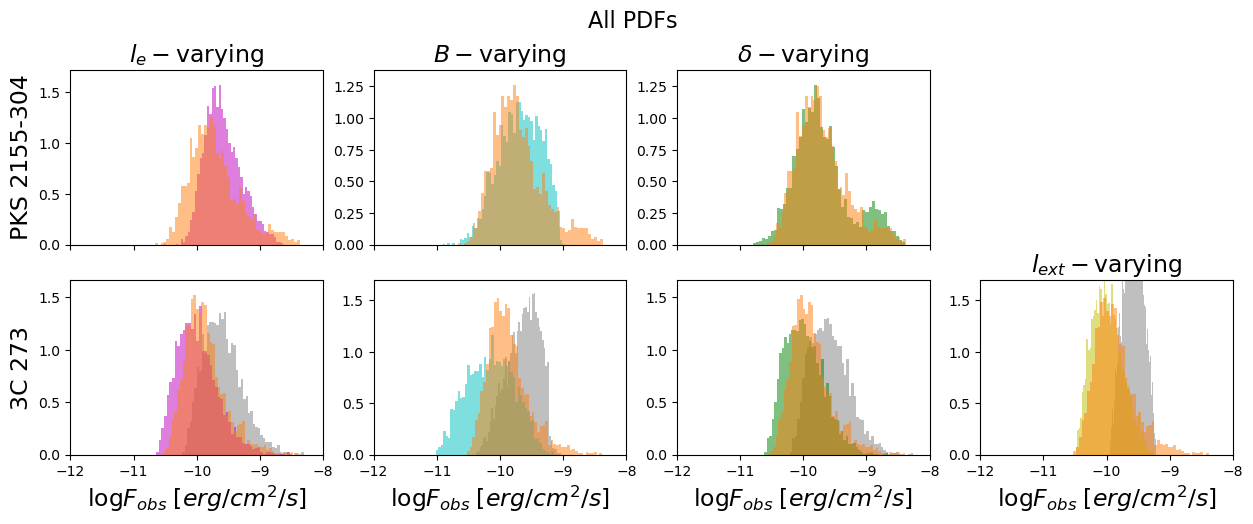}
\caption{PDFs of observed and simulated $\gamma$-ray fluxes in the $0.1-300$ GeV energy range for PKS~2155-304 (top panels) and 3C~273 (bottom panels). The grey histograms correspond to the model with $B_0=10$~G. The observed PDFs  (orange) are created using the 10 year-long daily-binned \fermi-LAT light curves. By definition, all histograms are normalized so that the enclosed areas are equal to 1. For PKS~2155-304, the best agreement between simulated and observed PDFs is found for a variable $\delta$. For 3C~273, the best agreement is found for the $l_{\rm e}$-varying simulation with $B_0=22$~G, while no simulation with $B_0=10$~G produces desirable results.}
\label{fig:allPDFs}
\end{figure*}
\subsection{PDFs of $\gamma$-ray fluxes}\label{sec:pdfs}
The first important test for our methodology (schematically shown in Figure~\ref{fig:inoutMK}) is the direct comparison of the flux PDFs of simulated and real \fermi-LAT $\gamma$-ray light curves. 

Results for the SSC scenario, which is used to model the SED of PKS~2155-304, are presented in the top panels of  Figure~\ref{fig:allPDFs}. For all simulations, the mean of the simulated and observed $\gamma$-ray flux PDFs is similar as expected; the time-dependent simulations started from a steady-state model that described well the average \fermi-LAT spectrum (see dashed black lines in Figure~\ref{fig:vFvs_pks}). Even though the 68\% percentiles of both PDFs in the $l_{\rm e}$-varying simulation are similar, we cannot reproduce the bimodal PDF of observed fluxes (see orange histogram). 
This is related to the power-law index $\sigma_{\gamma}$ (see equation~\ref{eq:normx}) that was taken to be constant throughout the simulation. Using analytical arguments, one can show that $\sigma_{\gamma} \approx 2$ in the case of synchrotron cooling (see e.g. \citet{Bloom96} and Appendix~\ref{sec:app}). This assumption breaks down, however, when SSC cooling of electrons becomes equally important to synchrotron cooling. Roughly speaking, this happens whenever the SSC ($\gamma$-ray) luminosity exceeds the synchrotron luminosity. In this regime, the appropriate power-law index to be used in the transformation should be $<2$ (see also Appendix~\ref{sec:app}). This would translate to larger variations of $l_{\rm e}$ around its mean value and to a broadening of the $\gamma$-ray flux PDF. Similarly, the $B$-varying simulations cannot capture the bimodality of the observed distribution of $\gamma$-ray fluxes. For the $B$-varying simulation, the value of the constant injected electron luminosity poses a hard upper limit to the $\gamma$-ray luminosity whenever particles are cooling mostly via SSC (see equation~(\ref{eq:ssc}) in Appendix~\ref{sec:app}). As a result, the PDF of simulated $\gamma$-ray fluxes shows a sharp cutoff at $\gtrsim10^{-9}$~erg cm$^{-2}$ s$^{-1}$. This finding is independent of the exact value for $\sigma_{\gamma}$. 
Finally, the best agreement between simulated and observed $\gamma$-ray flux PDFs for PKS~2155-304 is found for the $\delta$-varying simulation. Here, the imposed parameter variations do not change the relative importance of the synchrotron and ICS processes in particle cooling. As a result, the time-dependent SED retains its overall shape (see also panel c in Figure~\ref{fig:vFvs_pks}); any changes in the integrated $\gamma$-ray fluxes can be mapped almost directly to the variations of the input time series, hence of the synthetic light curves.

Results for the ECS scenario, which is used to describe the SED of 3C~273, are presented in the bottom panels of Figure~\ref{fig:allPDFs}. In all panels, we show results of simulations starting from two steady states that are characterized by different magnetic field strengths (see panels in the left hand side of Figure~\ref{fig:vFvs_3c}). The PDFs of all our simulations depend strongly on the adopted value of $B_0$. For example, the PDF for $B_0=10$~G is systematically shifted to higher $\gamma$-ray fluxes compared to the PDF for $B_0=22$~G in all simulations, demonstrating that the source is on average more Compton dominated in the former than in the latter case (compare left-hand and right-hand side panels in Figure~\ref{fig:vFvs_3c}). The systematic offset between the mean values of the simulated PDFs for the two values of $B_0$ stems from the choice of the steady-state model (compare dashed black lines in the \fermi-LAT energy range in the left and right panels of Figure~\ref{fig:vFvs_3c}). When compared to the observed PDF of $\gamma$-ray fluxes, we find a qualitative agreement with the simulations with variable $l_{\rm e}$ and $\delta$ (and to a lesser degree $l_{\rm ext}$) 
for $B_0=22$~G. No simulation with $B_0=10$~G can produce PDFs close to the observed one. Hence, we will not consider any further the  steady-state model with $B_0=10$~G.

To summarize the findings of this section, we present in Table~\ref{tab:pdfs} the mean, median, and 68\% percentile of $\gamma$-ray fluxes (in logarithm) of all histograms in Figure~\ref{fig:allPDFs}. In general, all simulations (except those of varying $B$ for both sources and $l_{\rm e}$ for PKS~2155-304) yield similar PDFs to the observed ones. The inability of the $B$-varying simulations to capture the shape of the observed $\gamma$-ray PDFs is not a matter of the chosen $\sigma_{\gamma}$ value, but of its assumed time independence. In the $B$-varying simulations of both sources (and to a lesser extent in the $l_{\rm e}-$varying simulation of PKS~2155-304), the relative efficiency of synchrotron and Compton cooling processes changes throughout the course of the simulation. As a result, the dependence of the $\gamma$-ray flux on the model parameters changes in time (see also Appendix~\ref{sec:app}). Thus, there is no universal $\sigma_{\gamma}$ value that can be used while making the transformation shown in equation~(\ref{eq:normx}). 

\begin{table}
    \centering
    \caption{Mean, median, and 68\% Percentile of $\gamma$-ray fluxes (in logarithm) from the PDFs shown in Figure~\ref{fig:allPDFs}.}\begin{threeparttable}
    \begin{tabular}{cccc}
    \hline
     & Mean & Median &  68\% Percentile \\ 
    \hline
     \multicolumn{4}{c}{PKS~2155-304} \\
     obs & -9.7 & -9.8 & 0.6 \\
     $l_{\rm e}$ & -9.6 & -9.7 & 0.7 \\
     $B$ & -9.7 &  -9.7 &  0.7 \\
     $\delta$ & -9.7 &  -9.8 & 0.85 \\ 
     \hline 
    \multicolumn{4}{c}{3C~273}  \\
     obs &  -9.7 & -9.8  & 0.8 \\
     $l_{\rm e}$ &  -10.0  (-9.6) & -10.0 (-9.6) & 0.65 (0.6) \\
     $B$ &-10.1 (-9.6) &  -10.1 (-9.6) & 0.8 (0.55)
     \\
     $\delta$ & -10.0 (-9.6) &  -10.0 (-9.6) & 0.6 (0.6) \\ 
     $l_{\rm ext}$ & -10.0 (-9.6) &  -10.0 (-9.6) & 0.5 (0.3)\\
     \hline
    \end{tabular}
    \begin{tablenotes}
    Note -- Values in parentheses correspond to the steady-state model with $B_0=10$~G.
\end{tablenotes}
    \end{threeparttable}
    \label{tab:pdfs}
\end{table}

\newpage
\begin{figure}
\centering
\includegraphics[width = 1.0\linewidth]{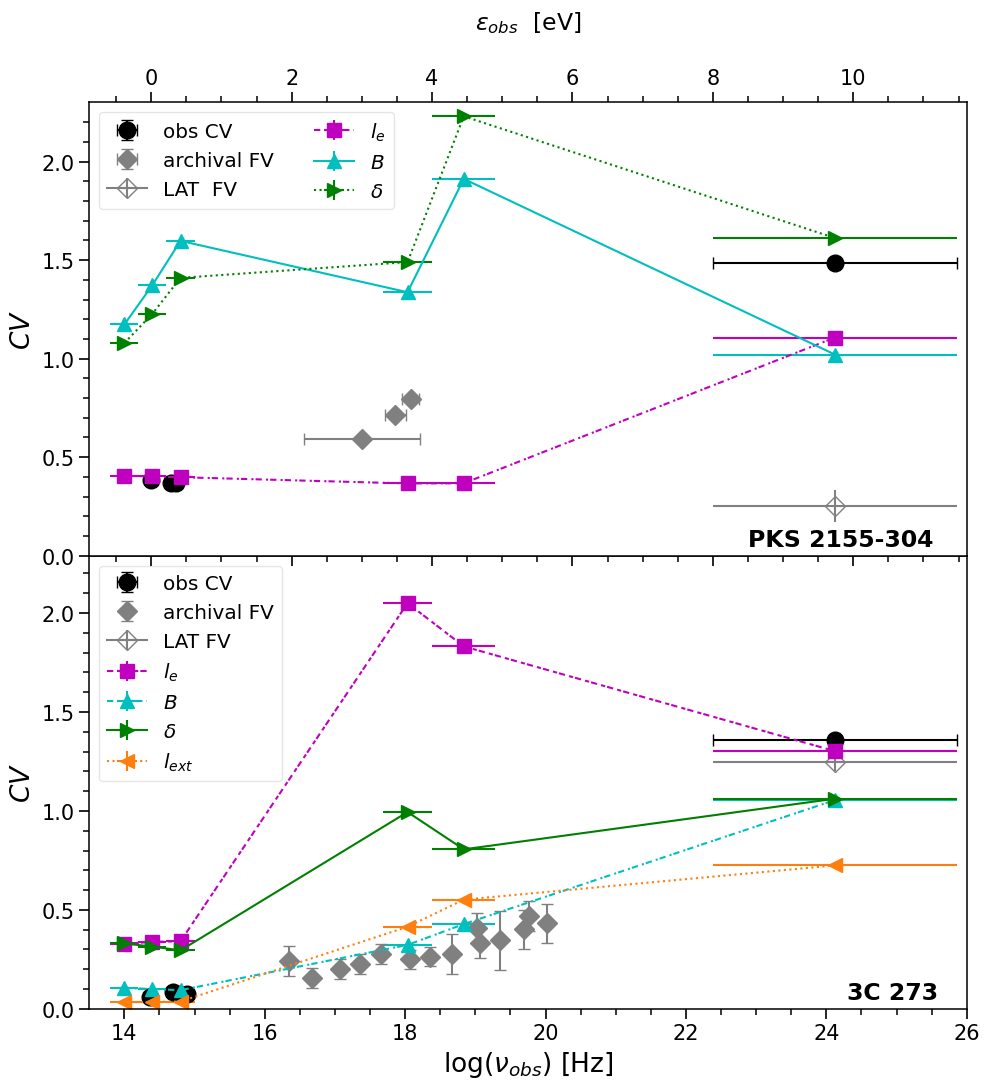}
\caption{Coefficient of variation (CV) as a function of frequency (coloured symbols) computed from our simulations for PKS~2155-304 (top panel) and 3C~273 (bottom panel). Black circles show the CVs in OIR and $\gamma$-ray energies computed from the observed SMARTS and \fermi-LAT light curves. The fractional variability (FV) value from the LAT light curve is also plotted for comparison (open diamond). FVs from archival observations in ultraviolet and X-rays (grey diamonds) are adopted from \citet{Chevalier} and \citet{Soldi} for PKS~2155-304 and 3C~273, respectively. These can be used as proxy of the CVs, as explained above.
}
\label{fig:allCVs}
\end{figure}

\subsection{Coefficient of variation}
To quantify the intensity of the variability in the simulated light curves we use the coefficient of variation  (CV, also known as relative standard deviation). This is defined as
\eq{CV}{
CV =\frac{\sqrt{\sigma^2}}{\langle f\rangle},
}
where $\langle f\rangle$ is the mean and $\sigma^2=\sum^N_{\rm i=1}(f_{\rm i}-\langle f \rangle)^2/(N-1)$ is the variance of the light curve consisting of $N$ data points with fluxes $f_{\rm i}$. 

CV relates to fractional variability (FV), a quantity commonly used in the analysis of observed time series \citep[for blazars, see e.g.][and references therein]{glx1}, as follows
\eq{FV}{ FV =\sqrt{CV^2 - \frac{\langle\sigma_{\rm err}^2\rangle}{\langle f\rangle^2}},}
where $\langle\sigma_{\rm err}^2\rangle =  \sum_{\rm i=1}^N \sigma_{\rm err,i}^2$ is the mean square error of the flux measurement uncertainties $\sigma_{\rm err,i}$. 
In general, CVs are found to follow the trend of calculated FVs \citep[see e.g.][]{Vaughan03} while we expect $FV \approx CV$ whenever the mean square error of flux measurements is much smaller than the intrinsic variance of the time series. This is usually the case for the observed light curves at OIR wavelengths and X-rays, but in $\gamma$-rays the statistical errors can be a significant source of variability. 

We compute the CVs of the daily binned 10-year long theoretical light curves for both sources in three energy ranges, namely OIR ($K, J, B$ filters), X-rays ($2-10$~keV and $10-80$~keV), and $\gamma$-rays ($0.1-300$~GeV). 
Our results for PKS~2155-304 and 3C~273 are presented in the top and bottom panels, respectively, of Figure~\ref{fig:allCVs} (coloured symbols). CVs computed from the observed 10-year SMARTS and \fermi-LAT light curves (filled blacks circles) are also included in the plots, as they can be directly compared to the values from our simulations. To illustrate the difference between CVs and FVs, we show the FV in $\gamma$-rays using the full \fermi-LAT light curve (open black diamond). As explained above, we find $CV > FV$ with a larger difference for PKS~2155-304. We have also checked that $CV \approx FV$ in OIR (not explicitly shown in the figure). For comparison reasons, we also show   observed FVs in X-rays from archival data in (filled grey symbols) adopted from \cite{Chevalier}  and \cite{Soldi} for PKS~2155-304 and 3C~273, respectively.

Variations of $l_{\rm e}$ (magenta squares) or $\delta$ (green right-pointing triangles) lead to much stronger X-ray variability than observed in 3C~273, thus suggesting that the variation of either one of these parameters alone is not plausible. This conclusion also applies to the broader class of FSRQ sources, when described by an one-zone ECS model. On the contrary, the simulations of varying $B$ (cyan hexagons), and on a lesser degree of $l_{\rm ext}$ (orange up-pointing triangles) 
produce variability in OIR and X-rays closer to the observed values, 
while reproducing the observed trend of stronger variability at higher energies. However, both simulations under-predict the strength of variability in $\gamma$-rays (see also Figure~\ref{fig:allPDFs}). 
 
The dependence of the CV on frequency differs significantly between the SSC and ECS scenarios, even when the same model parameter is allowed to vary (compare top and bottom panels in Figure~\ref{fig:allCVs}). For instance, the $B$-varying and $\delta$-varying simulations of PKS~2155-304 produce much more variable light curves than the observed ones in OIR and X-rays, but yield similar trends as the observed one. Simulations with variable $l_{\rm e}$ match the observed CVs at OIR, but underestimate the intensity of variability in the X-rays and $\gamma$-rays. The difference in the strength of $\gamma$-ray variability for the $l_{\rm e}$-varying simulation is also reflected to the shape of $\gamma$-ray flux PDFs (see top left panel in Figure~\ref{fig:allPDFs}). As explained in the previous section, this discrepancy could be alleviated if we allowed for small variations of the power-law index in equation~\ref{eq:normx} around the adopted value of $\sigma_\gamma=2$. The $l_{\rm e}$-varying simulation, however, cannot explain the observed trend of the CV versus frequency across the X-ray energy band. These findings suggest that changes of the maximum electron energy with time in combination with variations in $l_{\rm e}$ are likely needed for explaining the multi-wavelength observed variability. Interestingly, \citet{Chevalier}, using a different approach than ours, demonstrated that the variations in the cutoff Lorentz factor of the electron distribution are sufficient for explaining the observed high X-ray FV of PKS~2155-304. 

\begin{figure*} 
\centering
\includegraphics[width = 1.0 \textwidth]{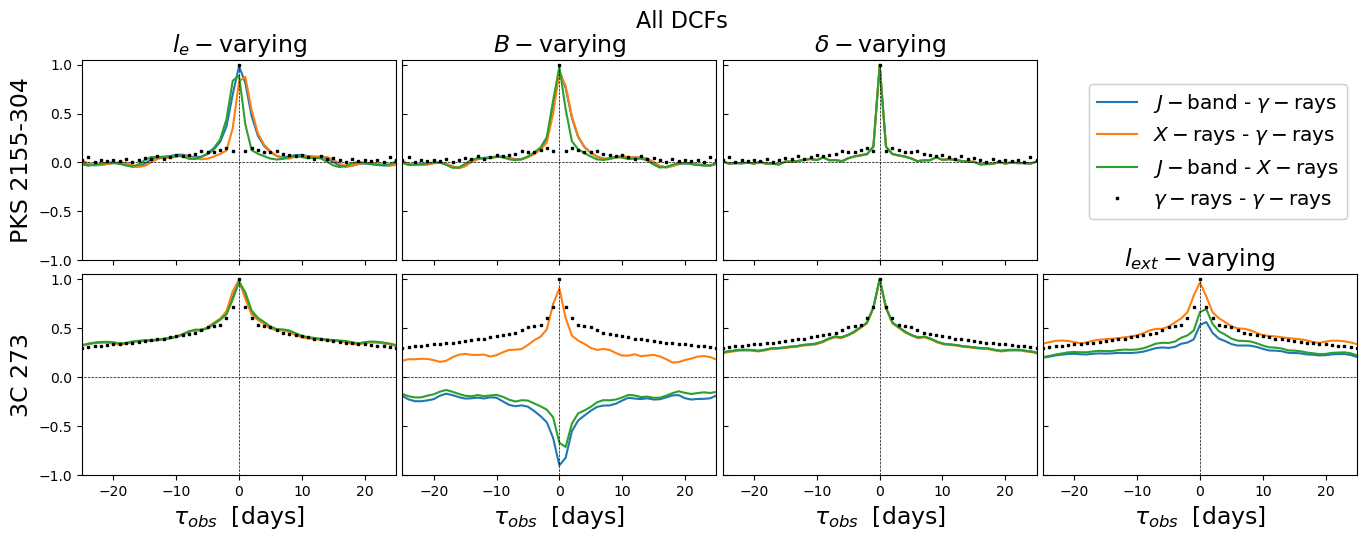}
\caption{Discrete Correlation Functions (DCFs) of the daily binned simulated light curves in the $J$-band, $2-10\;{\rm keV}$ X-rays  and $0.1-300$~GeV $\gamma$-rays for PKS~2155-304 (top panels) and 3C~273 (bottom panels). 
A positive time lag, $\tau_{\rm obs}$, between bands $a$ and $b$ means that the light curve in band $b$ lags behind the light curve in band $a$ by that amount of time. The DCF curves for each source are structured similarly to the auto-correlation function of the input \fermi-LAT light curve (shown in black symbols).}
\label{fig:allDCFs}
\end{figure*}

\subsection{Discrete Correlation Function}
To investigate the correlation and possible time lags between the simulated fluxes in OIR, X-rays, and $\gamma$-rays, we use the Discrete Correlation Function (DCF). For two discrete data series with the same length, $a_{\rm i}$ and $b_{\rm j}$, we compute first the unbinned discrete correlations for all pairs, $(a_{\rm i},b_{\rm j})$ \citep{DCF}
    \eq{UDCF}{UDCF_{\rm ij}=\frac{(a_{\rm i}-\langle a\rangle)(b_{\rm j}-\langle b \rangle)}{\sigma_{\rm a} \sigma_{\rm b}},}
where $\langle a \rangle$, $\langle b \rangle$  are the averages of $a_{\rm i}$ and $b_{\rm j}$ and $\sigma_{\rm a}$, $\sigma_{\rm b}$ are the respective standard deviations. Averaging the above expression for $M$ pairs of the time series with $\tau-\Delta \tau/2 <\Delta t_{\rm ij} <\tau+\Delta \tau/2$, where $\Delta t_{\rm ij}=t_{\rm i}-t_{\rm j} $, we calculated the DCF for time lag $\tau$,
    \eq{DCF}{DCF(\tau)=\frac{1}{M}\sum_{\rm i,j} UDCF_{\rm ij}.}
For two evenly binned time series with time resolution equal to $\Delta \tau$, the above expression yields the numerical approximation of the correlation function of the two time series.

The DCFs computed for three pairs of 10 year-long light curves ($J$ band versus $0.1-300$ GeV $\gamma$-rays, $2-10$ keV X-rays versus $J$ band, and $0.1-300$ GeV $\gamma$-rays versus $2-10$ keV X-rays) from our time-dependent simulations of PKS~2155-304 and 3C~273 are shown in Figure~\ref{fig:allDCFs}. When computing the DCFs for the simulations with varying $l_{\rm ext}$, we consider both the variable non-thermal jet emission in the $J$-band, but also the time-varying external radiation field. 
All simulated light curves are daily binned (for details, see Section~\ref{sec:method}). Results are only shown for $|{\tau}| < 25$~days, because of the lack of interesting features on
longer time lags. 

In the SSC scenario a strong positive correlation at zero time lag is found for all time dependencies we explored. Simulations with variable $l_{\rm e}$ and $B$ predict wider DCFs around zero lag than the DCF of the $\delta$-varying simulation. The broadening of the DCFs is related to the variable (intra-day) cooling timescale of radiating particles caused by changes in $B$ or $l_{\rm e}$, which are not present in the simulations of variable $\delta$. These results are not supported by the data, since no correlation between the observed long-term $J$-band and GeV $\gamma$-ray light curves of PKS~2155-304 on timescales $>1$~day was found \citep[][]{Bonning12,AtomHess,Yoshida2021}. If we ignored the model predictions for $|\tau|<1$~day, which cannot be probed by the daily-binned $\gamma$-ray observations, then only the predictions of the $\delta$-varying simulations are consistent with the data.

In the ECS scenario the DCFs are still symmetric around zero time lag, but broader compared to their respective DCFs for the SSC scenario. The shape of the DCFs is, however, not related to the emission scenario (i.e., SSC versus ECS) but with the timing properties of the light curve pair. To better illustrate this, we show the auto-correlation function (ACF) of the real $\gamma$-ray light curve of each source (black symbols). In general, the DCFs from the simulated light curves have shapes that are similar to the shape of the $\gamma$-ray ACF.

The ECS simulations exhibit a richer behaviour in the correlations found between the fluxes at different energy bands. For instance, the $B$-varying simulation yields an anti-correlation between the OIR and $\gamma$-ray emission from the jet, because of the varying relative importance of the synchrotron and Compton processes. A decrease in the magnetic field strength would reduce the synchrotron power and the OIR flux, while channeling more power to the ECS process and increasing the $\gamma$-ray flux respectively (see e.g. equations(\ref{eq:ecs-opt}) and ~(\ref{eq:ecs}) in Appendix~\ref{sec:app}).

An anti-correlation between the OIR and $\gamma$-ray fluxes is also found for the $l_{\rm ext}$-varying simulations when only the non-thermal fluxes (jet component) are considered (not explicitly shown in the figure). Higher $l_{\rm ext}$ values can lead to stronger cooling of electrons via ECS, thus reducing the power radiated via synchrotron in the OIR band. However, the $J$-band is dominated frequently by the time-dependent external photon field, which in turn correlates with $\gamma$-rays, eventually ``hiding'' the OIR-$\gamma$-ray anti-correlation from the jet. 

The correlation of the X-ray flux with the OIR and $\gamma$-ray fluxes in the ECS scenario is also affected by the relative importance of the SSC and ECS processes. For instance, in the $B$-varying simulation, the $2-10$~keV X-ray fluxes correlate most of the time with the flux in the $J$-band, hence the positive DCF values. The peak amplitude of the DCF curve is, however, slightly lower compared to that of other simulations, because there are time intervals where the X-rays correlate with $\gamma$-rays instead.
Both cases can be understood by the interplay of synchrotron or/and SSC (correlation; high values of magnetic field) and ECS (anti-correlation; low values of magnetic field) processes contributing to the intermediate energy bands of the SED \citep[see also][]{glx2}.

\begin{figure*}
\centering
\includegraphics[width=1.0\linewidth]{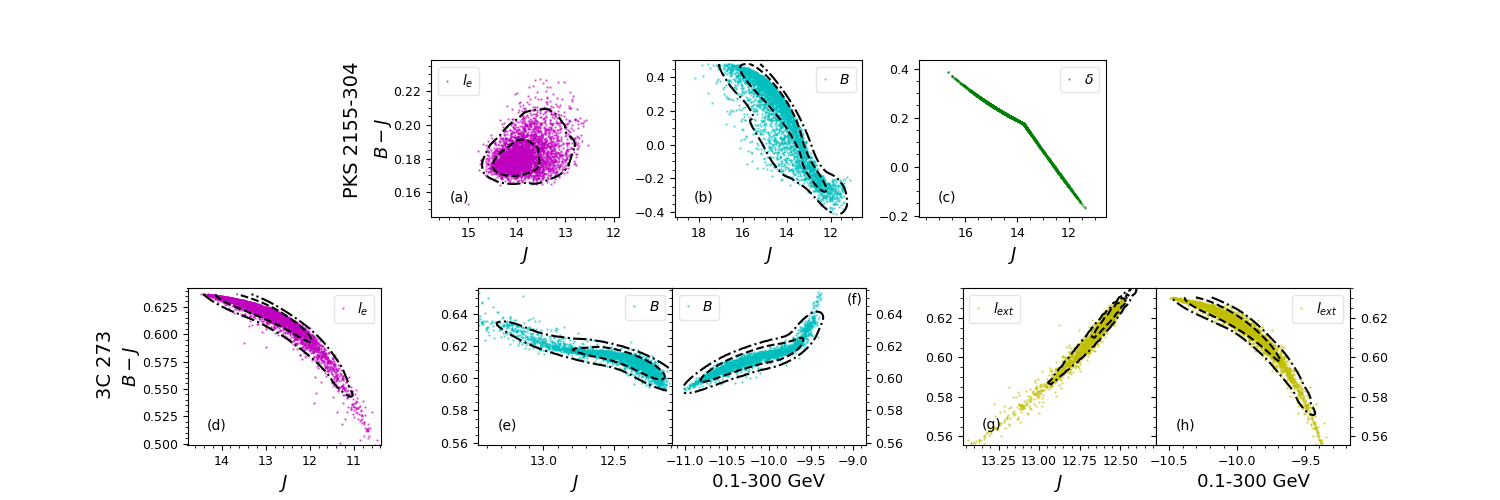}
\caption{Model-predicted $B-J$ colour plotted against the $J$-band magnitude for PKS~2155-304 and 3C~273. Results are obtained from the 10 year-long simulations of varying $l_{\rm e}$ (left), $B$ (center),  $\delta$ (top right), and $l_{\rm ext}$ (bottom right).
For two simulations of 3C~273 that produce an anti-correlation between OIR and $\gamma$-ray fluxes, the $B-J$ colour versus the $0.1-300$~GeV $\gamma$-ray flux is also plotted on the right of the respective colour-magnitude diagram. All other colour-$\gamma$-ray flux diagrams show similar trends as the $B-J$ versus $J$ plots and are not shown here.
Dashed (dotted-dashed) lines define the $1{\rm \sigma}$ ($2{\rm \sigma}$) standard deviation assuming a Gaussian kernel for the number density of states. No contours are plotted for the $\delta$-varying simulation of PKS~2155-304, because the distribution of points is practically one-dimensional. Note the different axis scales used in different panels. 
}
\label{fig:colors}
\end{figure*}

\begin{table*}
\centering
\caption{Summary of results from a qualitative comparison of simulations with observations using various diagnostics.}
\begin{threeparttable}
\begin{tabular}{|c|c||ccccc cc|}
\hline
    \multicolumn{2}{|c||}{}   & $\gamma$-ray PDFs &  
    \multicolumn{3}{c}{FV/CV} &  DCF ($J$-band vs. $\gamma$-rays)   & $B-J$ vs. $J$-band & $B-J$ vs. $\gamma$-rays \\
   \multicolumn{2}{|c||}{} &  & OIR & X-rays & $\gamma$-rays &  &  &  \\ 
\multicolumn{2}{|c||}{} & (1) & \multicolumn{3}{c}{(2)} & (3) & (4) & (5)  \\
   
   \hline
\multirow{3}{*}{\small{PKS 2155-304}} & $l_{\rm e}$  & \qmark   & \cmark & $\searrow$ & $\searrow$ &  \xmark & \cmark & \cmark \\
                                   & $B$  & \qmark  &  $\nearrow$  & $\nearrow$ & $\searrow$ & \xmark &  \xmark  & \xmark \\
                                  & $\delta$  & \cmark  & $\nearrow$ & $\nearrow$& \cmark & \qmark & \xmark & \xmark  \\
\hline
\multirow{4}{*}{\small{3C 273}} &$l_{\rm e}$    & \cmark & $\nearrow$  & $\nearrow$ &  \cmark & \xmark & \xmark & \xmark \\
                                     &$B$  & \xmark & \cmark & \cmark & $\searrow$  & \xmark & \qmark & \cmark \\
                                    &$\delta$  & \cmark & $\nearrow$ & $\searrow$   & $\searrow$  & \xmark & \xmark  & \xmark \\
                                    &$l_{\rm ext}$  & \cmark & \cmark   & \cmark & $\searrow$  & \qmark & \cmark & \xmark \\
\hline
\end{tabular}
\begin{tablenotes}
(1) The shape of the probability density function of the $\gamma$-ray light curve (see Figure~\ref{fig:allPDFs}).\\
(2) The shape of the multi-frequency coefficient of variation compared to the observed trends (see Figure \ref{fig:allCVs}).\\
(3) Shape of the DCF curves 
(see blue lines in Figure~\ref{fig:allDCFs} of this paper and Figure~4 of \citet{Bonning12} and \citet{Yoshida2021}).\\
(4) The shape of the $B-J$ colour versus $J$ magnitude colour diagram compared with observations (see Figure~\ref{fig:colors} of this paper and Figures~6 and 7 of \citet{Safna2020}). \\
(5) The shape of the $B-J$ colour versus  $\gamma-$ray flux colour diagram compared with observations (see Figure~\ref{fig:colors} of this paper and Figure~6 of \citet{Yoshida2021}).\\
\cmark (\xmark)  Property of the simulated light curve(s) is similar to (inconsistent with) the relevant property from  observations. \\
\qmark\, Controversial interpretation of the property under examination in the simulated results.\\
$\searrow$ ($\nearrow$)\, The model under-predicts (over-predicts) the observed CV/FV value. \\
\end{tablenotes}
\end{threeparttable}
\label{tab:summary}
\end{table*}

\subsection{Colour-Magnitude Diagrams}
Colour changes as a function of brightness in optical and infrared wavelengths is another diagnostic of blazar variability usually discussed in the literature \citep[see e.g.][]{Isler17}. In Figure~\ref{fig:colors} we present the $B-J$ versus $J$-band magnitude diagrams for both sources calculated using the 10 year-long near-IR simulated light curves for some of the time-dependencies studied so far. Moreover, the $B-J$ versus $0.1-300$~GeV diagrams of some cases that are worth discussing are also shown. The colour versus $\gamma-$ray flux diagrams which are not shown in the figure exhibit similar trends as those of the respective $B-J$ versus $J$ diagrams. 

The colour-magnitude diagram for the $\delta$-varying simulation of 3C~273 is similar to the one of PKS~2155-304 and is therefore not shown. The shape and range of colour variations produced by each type of time dependence can be used as a diagnostic of blazar variability. Changes of the same model parameter (here, $l_{\rm e}$ and $B$) can lead to very different colour variabilities within the SSC and ECS scenarios. We find no correlation between colour and magnitude for the $l_{\rm e}$-varying simulations of PKS~2155-304, in agreement with the analysis of long-term SMARTS observations presented by \citet{Safna2020}. A lack of correlation between intensity and colour in OIR filters 
($R$ and $H$) was also found by analyzing 7 years of data from the REM telescope \citep{Sandrinelli}. For certain 
shorter time periods, a bluer-when-brighter behaviour has been also reported for PKS~2155-304 \citep[see ][]{Paltani97}. The simulations of $B$ and $\delta$ are in qualitative agreement with this trend.

For 3C~273, a weak\footnote{For points lying within the $2{\rm \sigma}$ contours (dashed lines), the colour is approximately constant for all values of the $J$ magnitude, compared to the typical scale of colour 
changes ($\sim 1\,{\rm mag}$) in observations.} bluer-when-brighter behaviour is exhibited for the $l_{\rm e}$, $B$ and $\delta$ time-dependence, while the time-dependent external photon luminosity (see also Figure~\ref{fig:vFvs_3c}) produces a weak redder-when-brighter behaviour (see bottom right panel). In the latter case, we have assumed that the disk component remains constant in time. More realistic models that take into account the real spectral shape of the external radiation fields (e.g. BLR and disk), as well as the variability of the disk, may result in a different colour-brightness behaviour. Regarding the colour versus $\gamma$-ray flux diagrams of Figure~\ref{fig:colors}, a redder-when-brighter trend is exhibited for the $B$ time-dependence, while the colour remains constant and only slightly becomes bluer for high $\gamma$-ray fluxes in the case of a variable $l_{\rm ext}$. The latter behaviour is not anticipated based on the observed correlation between optical and $\gamma-$rays in the DCF plot (bottom right panel, Figure~\ref{fig:allDCFs}) and it highlights the underlying anti-correlation between the non-thermal OIR and $\gamma-$ray fluxes. 

The exact values of colour and magnitude are very sensitive to the choice of baseline parameter values for the power-law slope $p$ and minimum Lorentz factor $\gamma_{\min}$ of the electron distribution (since electrons close this energy emit in OIR frequencies). Given that the selection of the steady-state model was somewhat arbitrary (i.e., not aiming at the best-fit solution), systematic differences between the observed and simulated colours  and magnitudes are expected. Nevertheless, the $l_{\rm e}$-varying simulation of PKS~2155-304 and $l_{\rm ext}$-varying simulation of 3C~273 yield colour-magnitude trends that are in qualitative agreement with the observed ones~\citep{Bonning12, Isler14, Safna2020}.

The results presented in this section are summarized in Table~\ref{tab:summary}. We compared qualitatively the simulation results for each diagnostic listed in the table with observational findings from the literature. This table can serve as a roadmap when trying to determine the simulation(s) that best describe the multi-wavelength properties of each source. Simulations that fail to check most boxes are still useful, as they can guide us in the search of more complex models to describe the data (e.g. with two-parameter variations). For a discussion of our results, see Section~\ref{sec:discussion}.

\begin{figure*}
\centering
\includegraphics[width = 1.00\linewidth]{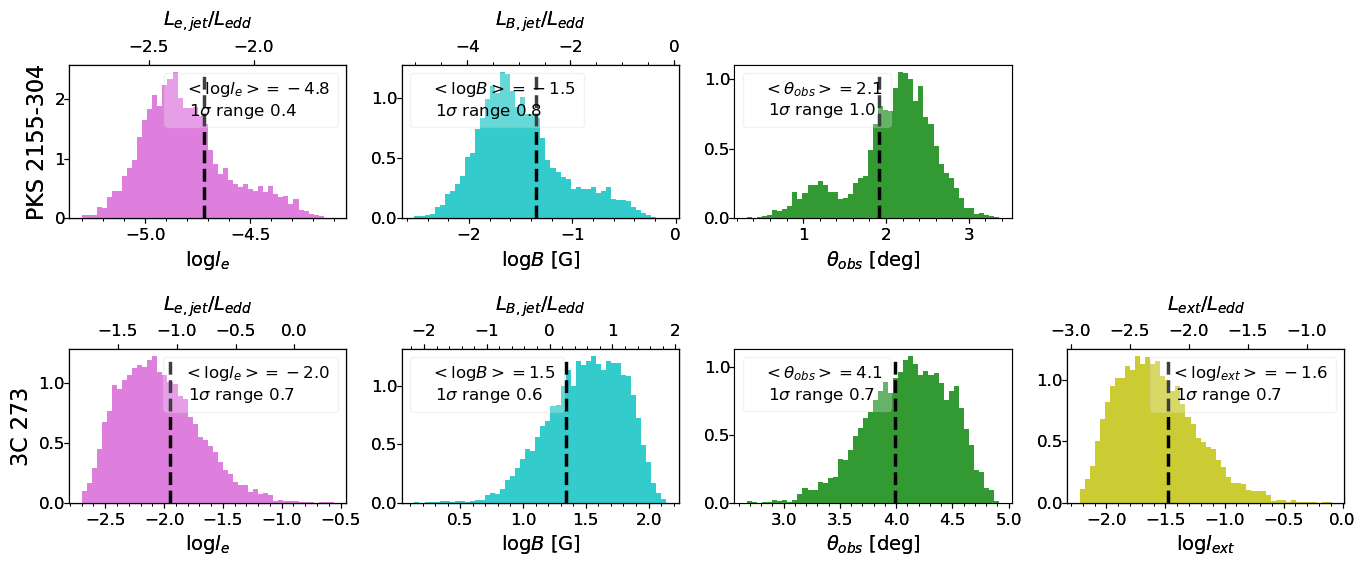}
\caption{Histograms of values for the variable model parameters obtained from long-term simulations for PKS~2155-304 (top row) and 3C~273 with $B_0=22$~G (bottom row). The top axis in each panels (except those for $\theta_{\rm obs}$) refers to the inferred jet luminosities normalized to the Eddington luminosity. 
Inset legends show the average  and the $1\sigma$ range of values (in logarithmic units) of each parameter distribution (for $\theta_{\rm obs}$ a linear scaled is used).
}
\label{fig:paramlims}
\end{figure*}

\subsection{Physical parameters and energetics}\label{sec:params}
We created time series for certain parameters of the one-zone leptonic model based on the observed properties of long-term $\gamma$-ray light curves. The distributions of the time-varying parameters used in our simulations of PKS~2155-304 and 3C~273 (for $B_0=22$~G) are presented respectively in the top and bottom panels of Figure~\ref{fig:paramlims}. All physical parameters, except the observer's angle $\theta_{\rm obs}$ (which is assumed to cause the changes in the Doppler factor), vary by a factor of 3 to 10 (see inset legends in panels of Figure~\ref{fig:paramlims}) within 10--100 days as measured in the blob frame. Our findings raise the question of what physical mechanisms (if any) could produce such variability in the physical conditions of the emitting region in blazar jets. We return to this point in Section~\ref{sec:discussion-phys}.

In Figure~\ref{fig:paramlims} we also show how the variation in $l_{\rm e}$ or $B$ translates to the relevant variable component of the jet power (see top axis), namely 
$L_{\rm e, jet}=(8\pi m_{\rm e} c^3)/(9 \sigma_{\rm T})\,R\,\Gamma^2\,l_{\rm e}$ and $L_{\rm B,\rm jet}=(2\pi m_{\rm e} c^3/\sigma_{\rm T})\,R\,\Gamma^2\,l_{\rm b}$. In the case of varying $l_{\rm ext}$, the relevant quantity to show is the bolometric external photon luminosity $L_{\rm ext}=(4\pi m_{\rm e} c^3/\sigma_{\rm T})\, R^{2}_{\rm ext}\,R^{-1}\,l_{\rm ext}$. Note that the jet power does not change when the Doppler factor variations are attributed to changes in $\theta_{\rm obs}$. All powers are normalized to the Eddington luminosity of each source, i.e., $L_{\rm edd}=1.26\times 10^{38}~(M/M_\odot)$~erg s$^{-1}$, where the black hole masses of 3C~273 and PKS~2155-304 are respectively $M=8.8\times 10^{8}\;M_\odot$ \citep{Peterson04} and $M=1.2\times 10^{8}\;M_\odot$ \citep{Gaur10}.

In the simulations with varying electron luminosity, the highest values of the $l_{\rm e}$ distribution correspond to $\gamma$-ray flares. In 3C~273, such flaring states are obtained for $L_{\rm e, jet} \lesssim L_{\rm Edd}$, while the jet luminosity remains always well below the Eddington limit for PKS~2155-304. These results are consistent with the proposal that FSRQs and BL Lac objects accrete at different rates~\citep{Baum95,Urry1995,Maraschi03,Ghis09}, assuming of course that the electron luminosity of the jet is related to the accretion power. The simulated magnetic luminosity of the jet, $L_{\rm B,jet}$, exceeds frequently the Eddington luminosity of 3C~273 by $\sim$one order of magnitude, while its steady-state value is close to $L_{\rm Edd}$. A more meaningful comparison would be that of the jet and accretion powers. Noting that the bolometric disk luminosity of 3C~273 is $L_{\rm disk}\sim L_{\rm Edd}$ \citep[][]{Malkan}, the accretion power is $\dot{M}c^2 =  L_{\rm disk}/\eta \approx 3\times10^{47}(0.3/\eta)$~erg s$^{-1}$, where $\eta$ is the radiative efficiency. The $l_{\rm e}$-varying simulations of 3C~273 have a jet power that is a fraction of the accretion power, while the $B$-varying simulations of 3C~273 predict states with $L_{\rm jet}\lesssim (3-5)\dot{M}c^2$ that are borderline consistent with the most efficient scenarios of jet formation \citep[see e.g.][]{Tchek11}.  Finally, in the $l_{\rm ext}$-varying simulations of 3C~273 we find that $L_{\rm ext} \ll L_{\rm Edd}$ for the whole duration of the simulation. These results are consistent with a BLR interpretation of the external photon field, as its luminosity is always a fraction of the near-Eddington accreting disk luminosity.

%
%
%
%
%
%
%
%
%
%
\section{Discussion}\label{sec:discussion}
In this section, we discuss the steady-state model selection and how this impacts our results. We also discuss factors that may affect the observed correlation patterns and the possibility of combining more than one time-dependent parameters of the model to reproduce blazar variability. We finally comment on the likely physical origins of the variability of the model parameters.

\subsection{The role of the steady-state model}
The selection of the steady-state parameters is a key step in our algorithm for producing the desired multi-wavelength variability patterns. In general, small changes in the steady-state parameter values will not qualitatively change the timing properties of the SED, unless the new selection of the steady state leads to frequent transits of the emitting electrons between cooling regimes in the course of the simulation. This has been demonstrated for 3C~273 using two steady-state models that differed in their magnetic field strength (see Table~\ref{tab:steady} and Figure~\ref{fig:vFvs_3c}).

The shape of the steady-state energy spectrum is another factor determining the flux variability in a certain energy band. Simulations of varying $l_{\rm e}$ and $l_{\rm ext}$ do not produce significant spectral changes and therefore are expected to yield similar multi-wavelength variability patterns for various time-average SEDs.
On the contrary, simulations of varying $B$ and $\delta$, which cause frequency shifts, are more sensitive in the selection of the average SED model. For instance, both such simulations of PKS~2155-304 predict extreme variability (large CV values, see left-hand side of Figure \ref{fig:allCVs}) in the $2-10$ keV energy band for the adopted value of $\gamma_{\max}$ (see middle and bottom panel of Figure~\ref{fig:vFvs_pks}). This excessive X-ray variability could have been reduced, if a different steady-state model with a higher $\gamma_{\max}$ value was selected. In this case, variations of $B$ or $\delta$ would rarely push the exponential cutoff of the synchrotron spectrum into the $2-10$~keV energy range, thus reducing the observed variability in that band.

The proximity of the colour bands to certain spectral features of the steady-state SED (e.g. synchrotron self-absorption frequency, minimum synchrotron frequency, or cooling break frequency) can be tuned by selecting certain set of parameters.These include parameters describing the injected electron population (e.g. $p$, $\gamma_{\min}$, and $\gamma_{\max}$) as well as $B$ and $l_{\rm ext}$.  As a result, the model-predicted colour and its temporal change are very sensitive on the selection of steady-state parameters that regulate the spectral changes of the SED. Moreover, the spectral shape of the BLR component (here modeled as a grey body of constant temperature) also affects the derived values of the $B-J$ colour in the simulations of varying $l_{\rm ext}$.
Fine-tuning of these parameters is necessary for rescaling the colour-magnitude diagrams to match the observed colours  and magnitudes of PKS~2155-304 and 3C~273. Note that this fine-tuning is not expected to change the patterns or the range of variations in the colour-magnitude diagrams.

The steady-state (baseline) parameter values of $R$ and $\delta$ determine the average dynamical timescale of the problem and the escape time of electrons, while the values of $B$ and $l_{\rm ext}$ determine the average synchrotron and ECS cooling timescales, respectively. Changes in either one of these timescales may result in changes of the cooling efficiency of electrons radiating at a fixed observing frequency. This may in turn affect the time lags between different energy bands. For instance, if a steady state was characterized by large values of $t_{\rm cr}$ (e.g. days), then time lags up to a few days could be found between energy bands, each probing different parts of the electron distribution (i.e., slow and fast cooling electrons). Such lags would produce asymmetric DCFs with a secondary bump at day-long lags.

\subsection{Shortcomings of the model}\label{sec:shortcomings}
We have produced time-dependent SEDs assuming that the blazar emission over long timescales arises from the same region of the jet and its temporal variability can be attributed to changes of one physical parameter. While this may be a simplification, it allowed us to isolate the effects that each parameter has on the long-term timing properties of blazar emission. Here, we present some shortcomings of our model based on our results for PKS~2155-304 and 3C~273. We also discuss how multi-parameter variations or multiple emission components could help to resolve them.

A strong correlation between the $J$-band (or X-ray) and $\gamma$-ray fluxes, with practically zero time lag, is found in all simulations of PKS~2155-304 (see top panels of Figure~\ref{fig:allDCFs}). This is a strong prediction of the one-zone SSC scenario with a one time-varying parameter. Observational studies of the long-term variability of PKS~2155-304 \citep[][]{Bonning12,AtomHess} found no correlation between the $J$-band and GeV $\gamma$-ray fluxes, in contrast to our model predictions. A more recent DCF analysis performed on 10 years of SMARTS and \fermi-LAT data supports these findings \citep{Yoshida2020, Yoshida2021}. Only the predictions of the $\delta$-varying simulations are consistent with the data, if $|\tau|<1$~day are ignored. Nonetheless, if the lack of observed correlation between the optical and GeV $\gamma$-ray fluxes persists on timescales shorter than $1$~day, it might suggest the presence of a second component that contributes to the OIR flux, and possibly to the X-ray flux, but has no significant contribution to the GeV flux. This could be an underlying dim variable disk-like component or a different region of the jet likely responsible also for the radio emission. This scenario would also account for correlated X-ray and $\gamma$-ray variability during flares \citep[][]{Coppi99,HESS_pks}, if only one of the two components dominates the emission during that state \citep[e.g.][]{Aharonian2009, Petro14}.  The hypothesis of two decoupled emission components was also found to explain equally well the X-ray and TeV $\gamma$-ray data during an outburst from Mkn 501 \citep{Kraw02}. These authors were able to reproduce the spectra and light curves of the outburst by assuming that the X-rays originated from a superposition of a soft quasi-steady component and a hard rapidly variable component, which could be attributed to either changes in $l_{\rm e}$ or $\delta$. It still remains to be shown if consideration of a steady emission component could help explaining the long-term flux variability properties of BL Lac objects, like PKS~2155-304.

The DCFs of the $J$-band and $\gamma$-ray light curves of 3C~273 computed within our model show a strong peak at zero time lag (see bottom panels of Figure~\ref{fig:allDCFs}). The $B$-varying simulation, in particular, predicts an anti-correlation of OIR and $\gamma$-ray fluxes, which has not been observed. Instead, a weak correlation at zero time lag was reported for 3C~273 by \citet{Bonning12} using 2 years of SMARTS and \fermi-LAT data \citep[see also][]{Soldi}. The weak peak at zero time lag was also recovered by a more recent DCF analysis performed on 10 years of SMARTS and \fermi-LAT data \citep{Yoshida2020, Yoshida2021}. This analysis revealed another peak at lag $\sim50$~days, which cannot be reproduced by any of our simulations. However, a weaker correlation at zero time lag than the one presented in Figure~\ref{fig:allDCFs} could be obtained if more than one model parameters were allowed to vary, thus bringing our model closer to the observed trends. This could be achieved for instance, if variations of the magnetic field strength, which produce anti-correlated OIR and $\gamma$-ray flux variability, were coupled to the variations of another physical parameter responsible for correlated flux variations (e.g. $l_{\rm e}$ or $\delta$). While a correlated variation of $l_{\rm e}$ and $B$ could mitigate this problem for the long-term variability properties of 3C~273, it might not apply to flaring FSRQs. For example, \citet{Bonnoli2011} found that an anti-correlation between the electron luminosity and the strength of the magnetic field is necessary for explaining the optical, X-ray and $\gamma$-ray flux changes during a bright $\gamma$-ray flare of FSRQ 3C~454.3. It is therefore possible that certain flares originate from a different region than the one responsible for the long-term and less variable blazar emission. Although this is mere speculation at this point, a systematic time-dependent modeling of flaring periods and long-term emission of selected FSRQs would be able to test this hypothesis.

Analysis of $\sim10$ year-long OIR and $\gamma$-ray observations of 3C~273 with SMARTS and \fermi-LAT, respectively, has shown a redder when $\gamma$-ray brighter behaviour \citep{Yoshida2020, Yoshida2021}. We find a similar behaviour only for the $B$-varying simulations of 3C~273 (bottom central
panel of Figure \ref{fig:colors}). However, this type of parameter variation leads to anti-correlated flux variability in OIR and $\gamma$-ray energies (see Figure \ref{fig:allDCFs}), in contrast to the observed correlation as previously discussed. Meanwhile, all other simulations reproduce the correlation and describe well the CV trend in the OIR and X-ray bands (see Figure \ref{fig:allCVs}) as well as the $B-J$ versus $J$ trend (Figure \ref{fig:colors}). The fact that none of our simulations can describe all the long-term variability properties of 3C~273 at once adds to the arguments against the single-parameter-varying scenario for this source and other FSRQs with similar behaviour.

Besides the time-varying BLR and jet components already considered in the simulations of 3C~273, the disk itself can be another source of variability in the optical/UV bands. Incorporation of these fluctuations in our simulations would require the use of a realistic time-dependent disk model \citep[``flickering'' and long-term variability, see ][]{Lyu97}. The need for a more realistic model of the thermal component in FSRQs is especially important for studying the variability properties of the optical emission, as this is often dominated by the non-jetted components.

Lastly, the fact that a model parameter, like the magnetic field strength, has to vary up to 2 orders of magnitude to fully account for the observed $\gamma$-ray variability  (see Figure \ref{fig:paramlims}) makes the assumption of a single-parameter variation questionable. For instance, one may expect that the conditions in the emission region are related to those in the region where particles are accelerated. If so, then the particle acceleration efficiency is not expected to remain constant while the magnetic field undergoes such dramatic changes in strength. In other words, $\gamma_{\max}$ and/or $p$ would also have to change together with $B$, leading to a more complex scenario for blazar variability. 
It is also worth mentioning that 
the tails of the distributions of the parameter values are obtained during the brightest $\gamma$-ray flares. In a two-component scenario, where $\gamma$-ray flares may originate from a region other than the one producing the smaller amplitude flux variability, the inferred parameter variations for the latter region would be less extreme than those found in this work.

\subsection{Physical implications}\label{sec:discussion-phys}
In Section~\ref{sec:params} we presented the variations of the model parameters used in this work, which were motivated by the observed properties of long-term LAT light curves of PKS~2155-304 and 3C~273. Here, we discuss if such variations are physically plausible.

The electron compactness, $l_{\rm e}$, is a dimensionless measure of the power injected into relativistic electrons via an acceleration process. A variable $l_{\rm e}$ can be associated with changes in the jet plasma density, assuming a constant acceleration efficiency at all times. Changes in the mass accretion rate onto the black hole is one mechanism capable of producing density irregularities in the jet. 
While the jet and mass accretion powers are correlated \citep[see e.g.][]{Ghisellini14}, a direct relation between changes in the mass accretion rate and the plasma density at distances far from the black hole has not been yet demonstrated (to the best of our knowledge). 
Time-dependent magnetohydrodynamic simulations of magnetically arrested accretion onto rotating black holes, which yields powerful jets, show that the accretion rate changes at most by a factor of 10 \citep{Tchek11}. This process could explain most of the states produced in our simulations (see e.g. $1\sigma$ values in inset legend of panels in the first column of Figure~\ref{fig:paramlims}). 
A density-dependent acceleration efficiency might be needed only for the  highest values of $l_{\rm e}$ that correspond to $\gamma$-ray flares.

In the case of the $B$-varying simulations, the values of the magnetic field span about one order of magnitude (see $1\sigma$ range and top axis respectively, in Figure~\ref{fig:paramlims}). A time variable magnetic field can be related to e.g. development of magnetohydrodynamic instabilities which may change the jet's local magnetic field \citep{Dong20} or compression by  shock waves \citep[][]{deHoffmann50,Marscher85,Keppens08,Summerlin11}. However, such processes cannot account for order-of-magnitude changes in the field strength implied by the observed $\gamma$-ray variability of both sources. Each of these mechanisms individually can not account for the variability of $B$ found in our simulations on all timescales, adding another disadvantage to the scenario of a time-varying magnetic field. However, a coupling of variations in the magnetic field strength with changes in the plasma density (as proposed for the $l_{\rm e}$ time-dependence) poses an interesting possibility that needs further exploration.

As shown in Figure~\ref{fig:paramlims}  (right panel), $l_{\rm ext}$ varies on average one order of magnitude around its mean value. Regarding the variations of the external photon compactness, these can be associated with changes in the properties of the BLR and/or disk. Changes in the luminosity of the clouds, for instance, linearly correlate with changes in the disk luminosity \citep[with a persistent timelag of hundreds of days, see][]{Zhang19}. In an observational study of 17 quasars, including 3C~273, \citet{Kaspi} showed that the optical continuum variability (at $510$ nm rest wavelength) is $\sim25\%-150\%$, while the BLR variability (as inferred from Balmer lines) is even smaller.
Therefore, a variable luminosity of the BLR clouds seems unlikely to account for the $l_{\rm ext}$ variations in our model. However, changes in the disk may influence the injected power into relativistic electrons, eventually introducing a two-parameter time-dependence in the one-zone model (i.e., $l_{\rm e}$ and $l_{\rm ext}$). The combined variations in these two parameters might require smaller changes in $l_{\rm ext}$.

Finally, the Doppler factor ranges between 15 and 60 for PKS~2155-304, and 7 to 19 for 3C~273. Such changes are not uncommon in studies of blazar flares. For instance, \citet{Kraw02} find similar range of values for the Doppler factor when modelling the X-ray and TeV flares of Mkn 501 within a SSC scenario. Given that the bulk Lorentz factor of the jet is unlikely to randomly exhibit such large changes, we assume that the Doppler factor variations are caused by changes in the observer's angle. For a constant bulk Lorentz factor $\Gamma=\Gamma_0$, we thus obtain the distribution of $\theta_{\rm obs}$ values presented in Figure~\ref{fig:paramlims}. The $1\sigma$ variation of the angle is within one degree from its average value. Time-dependence on the angle $\theta_{\rm obs}$ (as the only time-dependent parameter) suggests a rather chaotic trajectory of the blob, with alterations of the direction of relativistic motion produced on timescales limited by the characteristic length scale of the region of the jet the blob is considered to move in. In the case of random motions of the emitting region in the jet \citep[see e.g.][]{Ghis08,Giannios09,Narayan12,Biteau12,Wehrle16} we would expect multi-wavelength variations to be of statistical nature (i.e., white noise PSDs, Gaussian PDFs). Alternatively, multi-wavelength quasi-periodic oscillations (QPOs) would be expected, if periodic motion (e.g. helical) of the blob was assumed \citep{Sarkar2021}. The broken power-law PSDs of $\theta_{\rm obs}$ created using the \fermi-LAT light curves imply a more complex description of the blob motion.

%
%
%
%
%
%
%
%
%
\section{Conclusions}\label{sec:conclusions}
We performed a theoretical study of multi-wavelength blazar variability on long timescales using as test beds the BL Lac object PKS~2155-304 and the FSRQ 3C~273. For this purpose, we introduced time-dependence to four main parameters of the one-zone leptonic model. These are the injection electron luminosity $L_{\rm e}$ (which is related to the injection mechanism of relativistic particles in the blob), the magnetic field strength $B$, the external photon luminosity $L_{\rm ext}$ (which controls the cooling mechanism of particles inside the blob), and the Doppler factor $\delta$ (which can be related to the relativistic motion of the blob inside the jet). The time series for each model parameter were generated from the \fermi-LAT 10 year-long light curves using a power-law transformation (equation~\ref{eq:normx}) with an exponent that is physically motivated and reproduces the observed $\gamma$-ray PDFs.

For each time-varying parameter, we computed time-dependent SEDs of PKS~2155-304 and 3C~273 for a period of $\sim10$ years in the context of an SSC and an ECS scenario, respectively. Using various diagnostics, such as flux PDFs, DCFs and CVs, we checked the capability of each model to reproduce the observed blazar variability in three bands: $J$-band, X-rays ($2-10$ keV) and $\gamma$-rays ($0.1-300$ GeV). Besides the $\gamma$-ray flux variations, which are by construction described well by most models, none of the simulations with a single-varying parameter can account for all observed flux variability properties in OIR, X-rays, and $\gamma$-rays and colour trends. As shown in Table \ref{tab:summary}, among the simulations considered for PKS~2155-304 we find that the simulation with time-dependence on $l_{\rm e}$ checks most of the boxes, while for 3C~273 the $l_{\rm ext}$ time-dependence is capable of producing most of the observed properties.

We conclude that time-dependence of a single parameter of the one-zone leptonic model describes only partially the variability properties in all three bands (OIR, X-rays, $\gamma$-rays) on long timescales. The shortcomings of our most promising simulations, e.g. the mismatch of the simulated $\gamma$-ray flux-colour diagram ($l_{\rm ext}$ simulation of 3C~273) and inability of describing the observed variability in certain bands (e.g. very large X-ray CVs in $l_{\rm e}-$varying simulation of PKS~2155-204) suggest that at least two physical parameters have to vary simultaneously to explain the long-term variability across the electromagnetic spectrum. Given that each blazar seems to have a distinct ``personality'' (i.e., multi-wavelength behaviour in the time domain), our study scratches the surface of the blazar variability problem and motivates a wider investigation of the SMARTS blazar sample with the methods presented here.

\section*{Data availability}
The pipeline developed for performing the simulations and post-processing the code output files is called {\sc BlaVar} and is available on  \href{https://github.com/MPolkas/BlaVar}{GitHub}. 
All observational data used in this work are obtained from published papers, except for the daily-binned \fermi-LAT light curve of PKS~2155-304 that was provided to us by Dr.~M.~Meyer. \\

\textit{Note:}  \cite{Yoshida2021} report results of cross correlations of the \fermi-LAT $\gamma$-ray and SMARTS OIR light curves for bright eight blazars monitored in 2008-2017. Several of us are also co-authors of that paper, which is under internal review from the \fermi collaboration. Preliminary results have been presented in \cite{Yoshida2020}.

\section*{Acknowledgements}
The authors would like to thank the anonymous referee for a very constructive report that helped to significantly improve the manuscript.  The authors would like to thank M. Meyer for providing the daily-binned \fermi-LAT  light curve of PKS~2155-304. M.~Petropoulou would like to thank the Department of Astronomy at Yale University for its hospitality during which this project was conceived. 
GV acknowledges support by NASA Grant Number 80NSSC21K0213, 80NSSC20K0803 and 80NSSC20K1107.




\bibliographystyle{mnras}
\bibliography{blavar}

\begin{thebibliography}{}
\makeatletter
\relax
\def\mn@urlcharsother{\let\do\@makeother \do\$\do\&\do\#\do\^\do\_\do\%\do\~}
\def\mn@doi{\begingroup\mn@urlcharsother \@ifnextchar [ {\mn@doi@}
  {\mn@doi@[]}}
\def\mn@doi@[#1]#2{\def\@tempa{#1}\ifx\@tempa\@empty \href
  {http://dx.doi.org/#2} {doi:#2}\else \href {http://dx.doi.org/#2} {#1}\fi
  \endgroup}
\def\mn@eprint#1#2{\mn@eprint@#1:#2::\@nil}
\def\mn@eprint@arXiv#1{\href {http://arxiv.org/abs/#1} {{\tt arXiv:#1}}}
\def\mn@eprint@dblp#1{\href {http://dblp.uni-trier.de/rec/bibtex/#1.xml}
  {dblp:#1}}
\def\mn@eprint@#1:#2:#3:#4\@nil{\def\@tempa {#1}\def\@tempb {#2}\def\@tempc
  {#3}\ifx \@tempc \@empty \let \@tempc \@tempb \let \@tempb \@tempa \fi \ifx
  \@tempb \@empty \def\@tempb {arXiv}\fi \@ifundefined
  {mn@eprint@\@tempb}{\@tempb:\@tempc}{\expandafter \expandafter \csname
  mn@eprint@\@tempb\endcsname \expandafter{\@tempc}}}

\bibitem[\protect\citeauthoryear{{Acciari} et~al.,}{{Acciari}
  et~al.}{2020}]{Acciari}
{Acciari} V.~A.,  et~al., 2020, \mn@doi [\apjs] {10.3847/1538-4365/ab89b5},
  \href {https://ui.adsabs.harvard.edu/abs/2020ApJS..248...29A} {248, 29}

\bibitem[\protect\citeauthoryear{{Acero} et~al.,}{{Acero} et~al.}{2015}]{3FGL}
{Acero} F.,  et~al., 2015, \mn@doi [\apjs] {10.1088/0067-0049/218/2/23}, \href
  {https://ui.adsabs.harvard.edu/abs/2015ApJS..218...23A} {218, 23}

\bibitem[\protect\citeauthoryear{{Ackermann} et~al.,}{{Ackermann}
  et~al.}{2016}]{Ackermann2016}
{Ackermann} M.,  et~al., 2016, \mn@doi [\apjl] {10.3847/2041-8205/824/2/L20},
  \href {https://ui.adsabs.harvard.edu/abs/2016ApJ...824L..20A} {824, L20}

\bibitem[\protect\citeauthoryear{{Aharonian} et~al.,}{{Aharonian}
  et~al.}{2005}]{pkshess}
{Aharonian} F.,  et~al., 2005, \mn@doi [\aap] {10.1051/0004-6361:20041853},
  \href {https://ui.adsabs.harvard.edu/abs/2005A&A...430..865A} {430, 865}

\bibitem[\protect\citeauthoryear{{Aharonian} et~al.,}{{Aharonian}
  et~al.}{2007}]{Aharonian07}
{Aharonian} F.,  et~al., 2007, \mn@doi [\apjl] {10.1086/520635}, \href
  {https://ui.adsabs.harvard.edu/abs/2007ApJ...664L..71A} {664, L71}

\bibitem[\protect\citeauthoryear{{Aharonian} et~al.,}{{Aharonian}
  et~al.}{2009a}]{Aharonian2009}
{Aharonian} F.,  et~al., 2009a, \mn@doi [\aap] {10.1051/0004-6361/200912128},
  \href {https://ui.adsabs.harvard.edu/abs/2009A&A...502..749A} {502, 749}

\bibitem[\protect\citeauthoryear{{Aharonian} et~al.,}{{Aharonian}
  et~al.}{2009b}]{HESS_pks}
{Aharonian} F.,  et~al., 2009b, \mn@doi [\apjl] {10.1088/0004-637X/696/2/L150},
  \href {https://ui.adsabs.harvard.edu/abs/2009ApJ...696L.150A} {696, L150}

\bibitem[\protect\citeauthoryear{{Aharonian}, {Barkov}  \&
  {Khangulyan}}{{Aharonian} et~al.}{2017}]{Aharonian2017}
{Aharonian} F.~A.,  {Barkov} M.~V.,   {Khangulyan} D.,  2017, \mn@doi [\apj]
  {10.3847/1538-4357/aa7049}, \href
  {https://ui.adsabs.harvard.edu/abs/2017ApJ...841...61A} {841, 61}

\bibitem[\protect\citeauthoryear{{Asano} \& {Hayashida}}{{Asano} \&
  {Hayashida}}{2015}]{Asano2015}
{Asano} K.,  {Hayashida} M.,  2015, \mn@doi [\apjl]
  {10.1088/2041-8205/808/1/L18}, \href
  {https://ui.adsabs.harvard.edu/abs/2015ApJ...808L..18A} {808, L18}

\bibitem[\protect\citeauthoryear{{Atwood} et~al.,}{{Atwood} et~al.}{2009}]{LAT}
{Atwood} W.~B.,  et~al., 2009, \mn@doi [APJ] {10.1088/0004-637X/697/2/1071},
  \href {https://ui.adsabs.harvard.edu/abs/2009ApJ...697.1071A} {697, 1071}

\bibitem[\protect\citeauthoryear{{Baum}, {Zirbel}  \& {O'Dea}}{{Baum}
  et~al.}{1995}]{Baum95}
{Baum} S.~A.,  {Zirbel} E.~L.,   {O'Dea} C.~P.,  1995, \mn@doi [\apj]
  {10.1086/176202}, \href
  {https://ui.adsabs.harvard.edu/abs/1995ApJ...451...88B} {451, 88}

\bibitem[\protect\citeauthoryear{{Begelman}, {Blandford}  \& {Rees}}{{Begelman}
  et~al.}{1984}]{Begelman1984}
{Begelman} M.~C.,  {Blandford} R.~D.,   {Rees} M.~J.,  1984, \mn@doi [Reviews
  of Modern Physics] {10.1103/RevModPhys.56.255}, \href
  {https://ui.adsabs.harvard.edu/abs/1984RvMP...56..255B} {56, 255}

\bibitem[\protect\citeauthoryear{{Biteau} \& {Giebels}}{{Biteau} \&
  {Giebels}}{2012}]{Biteau12}
{Biteau} J.,  {Giebels} B.,  2012, \mn@doi [\aap]
  {10.1051/0004-6361/201220056}, \href
  {https://ui.adsabs.harvard.edu/abs/2012A&A...548A.123B} {548, A123}

\bibitem[\protect\citeauthoryear{{Bloom} \& {Marscher}}{{Bloom} \&
  {Marscher}}{1996}]{Bloom96}
{Bloom} S.~D.,  {Marscher} A.~P.,  1996, \mn@doi [\apj] {10.1086/177092}, \href
  {https://ui.adsabs.harvard.edu/abs/1996ApJ...461..657B} {461, 657}

\bibitem[\protect\citeauthoryear{{Blumenthal} \& {Gould}}{{Blumenthal} \&
  {Gould}}{1970}]{BG70}
{Blumenthal} G.~R.,  {Gould} R.~J.,  1970, \mn@doi [Reviews of Modern Physics]
  {10.1103/RevModPhys.42.237}, \href
  {https://ui.adsabs.harvard.edu/abs/1970RvMP...42..237B} {42, 237}

\bibitem[\protect\citeauthoryear{{Bonning} et~al.,}{{Bonning}
  et~al.}{2012}]{Bonning12}
{Bonning} E.,  et~al., 2012, \mn@doi [\apj] {10.1088/0004-637X/756/1/13}, \href
  {https://ui.adsabs.harvard.edu/abs/2012ApJ...756...13B} {756, 13}

\bibitem[\protect\citeauthoryear{{Bonnoli}, {Ghisellini}, {Foschini},
  {Tavecchio}  \& {Ghirlanda}}{{Bonnoli} et~al.}{2011}]{Bonnoli2011}
{Bonnoli} G.,  {Ghisellini} G.,  {Foschini} L.,  {Tavecchio} F.,   {Ghirlanda}
  G.,  2011, \mn@doi [\mnras] {10.1111/j.1365-2966.2010.17450.x}, \href
  {https://ui.adsabs.harvard.edu/abs/2011MNRAS.410..368B} {410, 368}

\bibitem[\protect\citeauthoryear{{B{\"o}ttcher}}{{B{\"o}ttcher}}{2019}]{Boettcher2019}
{B{\"o}ttcher} M.,  2019, \mn@doi [Galaxies] {10.3390/galaxies7010020}, \href
  {https://ui.adsabs.harvard.edu/abs/2019Galax...7...20B} {7, 20}

\bibitem[\protect\citeauthoryear{{B{\"o}ttcher} \& {Dermer}}{{B{\"o}ttcher} \&
  {Dermer}}{2010}]{Bottcher10}
{B{\"o}ttcher} M.,  {Dermer} C.~D.,  2010, \mn@doi [\apj]
  {10.1088/0004-637X/711/1/445}, \href
  {https://ui.adsabs.harvard.edu/abs/2010ApJ...711..445B} {711, 445}

\bibitem[\protect\citeauthoryear{{Chevalier}, {Sanchez}, {Serpico}, {Lenain}
  \& {Maurin}}{{Chevalier} et~al.}{2019}]{Chevalier}
{Chevalier} J.,  {Sanchez} D.~A.,  {Serpico} P.~D.,  {Lenain} J.~P.,   {Maurin}
  G.,  2019, \mn@doi [\mnras] {10.1093/mnras/stz027}, \href
  {https://ui.adsabs.harvard.edu/abs/2019MNRAS.484..749C} {484, 749}

\bibitem[\protect\citeauthoryear{{Chiaberge} \& {Ghisellini}}{{Chiaberge} \&
  {Ghisellini}}{1999}]{Chia99}
{Chiaberge} M.,  {Ghisellini} G.,  1999, \mn@doi [\mnras]
  {10.1046/j.1365-8711.1999.02538.x}, \href
  {https://ui.adsabs.harvard.edu/abs/1999MNRAS.306..551C} {306, 551}

\bibitem[\protect\citeauthoryear{{Cleary}, {Lawrence}, {Marshall}, {Hao}  \&
  {Meier}}{{Cleary} et~al.}{2007}]{cleary07}
{Cleary} K.,  {Lawrence} C.~R.,  {Marshall} J.~A.,  {Hao} L.,   {Meier} D.,
  2007, \mn@doi [\apj] {10.1086/511969}, \href
  {https://ui.adsabs.harvard.edu/abs/2007ApJ...660..117C} {660, 117}

\bibitem[\protect\citeauthoryear{{Coppi} \& {Aharonian}}{{Coppi} \&
  {Aharonian}}{1999}]{Coppi99}
{Coppi} P.~S.,  {Aharonian} F.~A.,  1999, \mn@doi [\apjl] {10.1086/312168},
  \href {https://ui.adsabs.harvard.edu/abs/1999ApJ...521L..33C} {521, L33}

\bibitem[\protect\citeauthoryear{{Dermer}}{{Dermer}}{1995}]{Dermer95}
{Dermer} C.~D.,  1995, \mn@doi [\apjl] {10.1086/187931}, \href
  {https://ui.adsabs.harvard.edu/abs/1995ApJ...446L..63D} {446, L63}

\bibitem[\protect\citeauthoryear{{Dermer} \& {Schlickeiser}}{{Dermer} \&
  {Schlickeiser}}{1993}]{Dermer93}
{Dermer} C.~D.,  {Schlickeiser} R.,  1993, \mn@doi [\apj] {10.1086/173251},
  \href {https://ui.adsabs.harvard.edu/abs/1993ApJ...416..458D} {416, 458}

\bibitem[\protect\citeauthoryear{{Dermer} \& {Schlickeiser}}{{Dermer} \&
  {Schlickeiser}}{2002}]{Dermer02}
{Dermer} C.~D.,  {Schlickeiser} R.,  2002, \mn@doi [\apj] {10.1086/341431},
  \href {https://ui.adsabs.harvard.edu/abs/2002ApJ...575..667D} {575, 667}

\bibitem[\protect\citeauthoryear{{Dermer}, {Schlickeiser}  \&
  {Mastichiadis}}{{Dermer} et~al.}{1992}]{Dermer92}
{Dermer} C.~D.,  {Schlickeiser} R.,   {Mastichiadis} A.,  1992, \aap, \href
  {https://ui.adsabs.harvard.edu/abs/1992A&A...256L..27D} {256, L27}

\bibitem[\protect\citeauthoryear{{Diltz} \& {B{\"o}ttcher}}{{Diltz} \&
  {B{\"o}ttcher}}{2014}]{Bottcher14}
{Diltz} C.,  {B{\"o}ttcher} M.,  2014, \mn@doi [Journal of High Energy
  Astrophysics] {10.1016/j.jheap.2014.04.001}, \href
  {https://ui.adsabs.harvard.edu/abs/2014JHEAp...1...63D} {1, 63}

\bibitem[\protect\citeauthoryear{{Dong}, {Zhang}  \& {Giannios}}{{Dong}
  et~al.}{2020}]{Dong20}
{Dong} L.,  {Zhang} H.,   {Giannios} D.,  2020, \mn@doi [\mnras]
  {10.1093/mnras/staa773}, \href
  {https://ui.adsabs.harvard.edu/abs/2020MNRAS.494.1817D} {494, 1817}

\bibitem[\protect\citeauthoryear{{Edelson} \& {Krolik}}{{Edelson} \&
  {Krolik}}{1988}]{DCF}
{Edelson} R.~A.,  {Krolik} J.~H.,  1988, \mn@doi [\apj] {10.1086/166773}, \href
  {https://ui.adsabs.harvard.edu/abs/1988ApJ...333..646E} {333, 646}

\bibitem[\protect\citeauthoryear{Emmanoulopoulos, McHardy  \&
  Papadakis}{Emmanoulopoulos et~al.}{2013}]{emmanou}
Emmanoulopoulos D.,  McHardy I.~M.,   Papadakis I.~E.,  2013, \mn@doi [Monthly
  Notices of the Royal Astronomical Society] {10.1093/mnras/stt764}, 433, 907

\bibitem[\protect\citeauthoryear{{Epitropakis} \& {Papadakis}}{{Epitropakis} \&
  {Papadakis}}{2016}]{epitropakis}
{Epitropakis} A.,  {Papadakis} I.~E.,  2016, \mn@doi [\aap]
  {10.1051/0004-6361/201527665}, \href
  {https://ui.adsabs.harvard.edu/abs/2016A&A...591A.113E} {591, A113}

\bibitem[\protect\citeauthoryear{{Gaur}, {Gupta}, {Lachowicz}  \&
  {Wiita}}{{Gaur} et~al.}{2010}]{Gaur10}
{Gaur} H.,  {Gupta} A.~C.,  {Lachowicz} P.,   {Wiita} P.~J.,  2010, \mn@doi
  [\apj] {10.1088/0004-637X/718/1/279}, \href
  {https://ui.adsabs.harvard.edu/abs/2010ApJ...718..279G} {718, 279}

\bibitem[\protect\citeauthoryear{{Ghisellini}}{{Ghisellini}}{2013}]{Ghisellini13}
{Ghisellini} G.,  2013, {Radiative Processes in High Energy Astrophysics}.
 Vol. 873, \mn@doi{10.1007/978-3-319-00612-3, }

\bibitem[\protect\citeauthoryear{{Ghisellini} \& {Tavecchio}}{{Ghisellini} \&
  {Tavecchio}}{2008}]{Ghis08}
{Ghisellini} G.,  {Tavecchio} F.,  2008, \mn@doi [\mnras]
  {10.1111/j.1745-3933.2008.00454.x}, \href
  {https://ui.adsabs.harvard.edu/abs/2008MNRAS.386L..28G} {386, L28}

\bibitem[\protect\citeauthoryear{{Ghisellini} \& {Tavecchio}}{{Ghisellini} \&
  {Tavecchio}}{2009}]{Ghisellini09}
{Ghisellini} G.,  {Tavecchio} F.,  2009, \mn@doi [\mnras]
  {10.1111/j.1365-2966.2009.15007.x}, \href
  {https://ui.adsabs.harvard.edu/abs/2009MNRAS.397..985G} {397, 985}

\bibitem[\protect\citeauthoryear{{Ghisellini}, {Padovani}, {Celotti}  \&
  {Maraschi}}{{Ghisellini} et~al.}{1993}]{Ghisellini93}
{Ghisellini} G.,  {Padovani} P.,  {Celotti} A.,   {Maraschi} L.,  1993, \mn@doi
  [\apj] {10.1086/172493}, \href
  {https://ui.adsabs.harvard.edu/abs/1993ApJ...407...65G} {407, 65}

\bibitem[\protect\citeauthoryear{{Ghisellini}, {Maraschi}  \&
  {Tavecchio}}{{Ghisellini} et~al.}{2009}]{Ghis09}
{Ghisellini} G.,  {Maraschi} L.,   {Tavecchio} F.,  2009, \mn@doi [\mnras]
  {10.1111/j.1745-3933.2009.00673.x}, \href
  {https://ui.adsabs.harvard.edu/abs/2009MNRAS.396L.105G} {396, L105}

\bibitem[\protect\citeauthoryear{{Ghisellini}, {Tavecchio}, {Maraschi},
  {Celotti}  \& {Sbarrato}}{{Ghisellini} et~al.}{2014}]{Ghisellini14}
{Ghisellini} G.,  {Tavecchio} F.,  {Maraschi} L.,  {Celotti} A.,   {Sbarrato}
  T.,  2014, \mn@doi [\nat] {10.1038/nature13856}, \href
  {https://ui.adsabs.harvard.edu/abs/2014Natur.515..376G} {515, 376}

\bibitem[\protect\citeauthoryear{{Giannios}, {Uzdensky}  \&
  {Begelman}}{{Giannios} et~al.}{2009}]{Giannios09}
{Giannios} D.,  {Uzdensky} D.~A.,   {Begelman} M.~C.,  2009, \mn@doi [\mnras]
  {10.1111/j.1745-3933.2009.00635.x}, \href
  {https://ui.adsabs.harvard.edu/abs/2009MNRAS.395L..29G} {395, L29}

\bibitem[\protect\citeauthoryear{{Giannios}, {Uzdensky}  \&
  {Begelman}}{{Giannios} et~al.}{2010}]{Giannios2010}
{Giannios} D.,  {Uzdensky} D.~A.,   {Begelman} M.~C.,  2010, \mn@doi [\mnras]
  {10.1111/j.1365-2966.2009.16045.x}, \href
  {https://ui.adsabs.harvard.edu/abs/2010MNRAS.402.1649G} {402, 1649}

\bibitem[\protect\citeauthoryear{{H.~E.~S.~S. Collaboration}
  et~al.,}{{H.~E.~S.~S. Collaboration} et~al.}{2014}]{AtomHess}
{H.~E.~S.~S. Collaboration} et~al., 2014, \mn@doi [\aap]
  {10.1051/0004-6361/201424142}, \href
  {https://ui.adsabs.harvard.edu/abs/2014A&A...571A..39H} {571, A39}

\bibitem[\protect\citeauthoryear{{H{\"o}nig} \& {Beckert}}{{H{\"o}nig} \&
  {Beckert}}{2007}]{honig07}
{H{\"o}nig} S.~F.,  {Beckert} T.,  2007, \mn@doi [\mnras]
  {10.1111/j.1365-2966.2007.12157.x}, \href
  {https://ui.adsabs.harvard.edu/abs/2007MNRAS.380.1172H} {380, 1172}

\bibitem[\protect\citeauthoryear{{Hovatta} et~al.,}{{Hovatta}
  et~al.}{2014}]{Hovatta2014}
{Hovatta} T.,  et~al., 2014, \mn@doi [\mnras] {10.1093/mnras/stt2494}, \href
  {https://ui.adsabs.harvard.edu/abs/2014MNRAS.439..690H} {439, 690}

\bibitem[\protect\citeauthoryear{{Hovatta} et~al.,}{{Hovatta}
  et~al.}{2015}]{Hovatta2015}
{Hovatta} T.,  et~al., 2015, \mn@doi [\mnras] {10.1093/mnras/stv220}, \href
  {https://ui.adsabs.harvard.edu/abs/2015MNRAS.448.3121H} {448, 3121}

\bibitem[\protect\citeauthoryear{{Isler}}{{Isler}}{2014}]{Isler14}
{Isler} J.~C.,  2014, PhD thesis, Yale University

\bibitem[\protect\citeauthoryear{{Isler}, {Urry}, {Coppi}, {Bailyn}, {Brady},
  {MacPherson}, {Buxton}  \& {Hasan}}{{Isler} et~al.}{2017}]{Isler17}
{Isler} J.~C.,  {Urry} C.~M.,  {Coppi} P.,  {Bailyn} C.,  {Brady} M.,
  {MacPherson} E.,  {Buxton} M.,   {Hasan} I.,  2017, \mn@doi [\apj]
  {10.3847/1538-4357/aa79fc}, \href
  {https://ui.adsabs.harvard.edu/abs/2017ApJ...844..107I} {844, 107}

\bibitem[\protect\citeauthoryear{{Jones}, {O'dell}  \& {Stein}}{{Jones}
  et~al.}{1974}]{Jones74}
{Jones} T.~W.,  {O'dell} S.~L.,   {Stein} W.~A.,  1974, \mn@doi [\apj]
  {10.1086/152724}, \href
  {https://ui.adsabs.harvard.edu/abs/1974ApJ...188..353J} {188, 353}

\bibitem[\protect\citeauthoryear{{Kardashev}}{{Kardashev}}{1962}]{Kardashev}
{Kardashev} N.~S.,  1962, \sovast, \href
  {https://ui.adsabs.harvard.edu/abs/1962SvA.....6..317K} {6, 317}

\bibitem[\protect\citeauthoryear{{Kaspi}, {Smith}, {Netzer}, {Maoz}, {Jannuzi}
  \& {Giveon}}{{Kaspi} et~al.}{2000}]{Kaspi}
{Kaspi} S.,  {Smith} P.~S.,  {Netzer} H.,  {Maoz} D.,  {Jannuzi} B.~T.,
  {Giveon} U.,  2000, \mn@doi [\apj] {10.1086/308704}, \href
  {https://ui.adsabs.harvard.edu/abs/2000ApJ...533..631K} {533, 631}

\bibitem[\protect\citeauthoryear{{Katarzy{\'n}ski} \&
  {Ghisellini}}{{Katarzy{\'n}ski} \& {Ghisellini}}{2007}]{Katarzynski07}
{Katarzy{\'n}ski} K.,  {Ghisellini} G.,  2007, \mn@doi [\aap]
  {10.1051/0004-6361:20066448}, \href
  {https://ui.adsabs.harvard.edu/abs/2007A&A...463..529K} {463, 529}

\bibitem[\protect\citeauthoryear{{Katarzy{\'n}ski}, {Ghisellini}, {Tavecchio},
  {Maraschi}, {Fossati}  \& {Mastichiadis}}{{Katarzy{\'n}ski}
  et~al.}{2005}]{Katarzynski2005}
{Katarzy{\'n}ski} K.,  {Ghisellini} G.,  {Tavecchio} F.,  {Maraschi} L.,
  {Fossati} G.,   {Mastichiadis} A.,  2005, \mn@doi [\aap]
  {10.1051/0004-6361:20041556}, \href
  {https://ui.adsabs.harvard.edu/abs/2005A&A...433..479K} {433, 479}

\bibitem[\protect\citeauthoryear{{Katarzy{\'n}ski}, {Ghisellini},
  {Mastichiadis}, {Tavecchio}  \& {Maraschi}}{{Katarzy{\'n}ski}
  et~al.}{2006}]{Katarzynski06}
{Katarzy{\'n}ski} K.,  {Ghisellini} G.,  {Mastichiadis} A.,  {Tavecchio} F.,
  {Maraschi} L.,  2006, \mn@doi [\aap] {10.1051/0004-6361:20054176}, \href
  {https://ui.adsabs.harvard.edu/abs/2006A&A...453...47K} {453, 47}

\bibitem[\protect\citeauthoryear{{Keppens}, {Meliani}, {van der Holst}  \&
  {Casse}}{{Keppens} et~al.}{2008}]{Keppens08}
{Keppens} R.,  {Meliani} Z.,  {van der Holst} B.,   {Casse} F.,  2008, \mn@doi
  [\aap] {10.1051/0004-6361:20079174}, \href
  {https://ui.adsabs.harvard.edu/abs/2008A&A...486..663K} {486, 663}

\bibitem[\protect\citeauthoryear{{Kishimoto}, {H{\"o}nig}, {Antonucci},
  {Barvainis}, {Kotani}, {Tristram}, {Weigelt}  \& {Levin}}{{Kishimoto}
  et~al.}{2011}]{Kishimoto11}
{Kishimoto} M.,  {H{\"o}nig} S.~F.,  {Antonucci} R.,  {Barvainis} R.,  {Kotani}
  T.,  {Tristram} K.~R.~W.,  {Weigelt} G.,   {Levin} K.,  2011, \mn@doi [\aap]
  {10.1051/0004-6361/201016054}, \href
  {https://ui.adsabs.harvard.edu/abs/2011A&A...527A.121K} {527, A121}

\bibitem[\protect\citeauthoryear{{Krawczynski}, {Coppi}  \&
  {Aharonian}}{{Krawczynski} et~al.}{2002}]{Kraw02}
{Krawczynski} H.,  {Coppi} P.~S.,   {Aharonian} F.,  2002, \mn@doi [\mnras]
  {10.1046/j.1365-8711.2002.05750.x}, \href
  {https://ui.adsabs.harvard.edu/abs/2002MNRAS.336..721K} {336, 721}

\bibitem[\protect\citeauthoryear{{Li}, {Zhang}, {Jin}, {Du}, {Cui}, {Liu}  \&
  {Wang}}{{Li} et~al.}{2020}]{2020ApJ...897...18L}
{Li} Y.-R.,  {Zhang} Z.-X.,  {Jin} C.,  {Du} P.,  {Cui} L.,  {Liu} X.,   {Wang}
  J.-M.,  2020, \mn@doi [\apj] {10.3847/1538-4357/ab95a3}, \href
  {https://ui.adsabs.harvard.edu/abs/2020ApJ...897...18L} {897, 18}

\bibitem[\protect\citeauthoryear{{Liodakis}, {Romani}, {Filippenko}, {Kocevski}
   \& {Zheng}}{{Liodakis} et~al.}{2019}]{Liodakis2019}
{Liodakis} I.,  {Romani} R.~W.,  {Filippenko} A.~V.,  {Kocevski} D.,   {Zheng}
  W.,  2019, \mn@doi [\apj] {10.3847/1538-4357/ab26b7}, \href
  {https://ui.adsabs.harvard.edu/abs/2019ApJ...880...32L} {880, 32}

\bibitem[\protect\citeauthoryear{{Lyubarskii}}{{Lyubarskii}}{1997}]{Lyu97}
{Lyubarskii} Y.~E.,  1997, \mn@doi [\mnras] {10.1093/mnras/292.3.679}, \href
  {https://ui.adsabs.harvard.edu/abs/1997MNRAS.292..679L} {292, 679}

\bibitem[\protect\citeauthoryear{{Malkan} \& {Sargent}}{{Malkan} \&
  {Sargent}}{1982}]{Malkan}
{Malkan} M.~A.,  {Sargent} W.~L.~W.,  1982, \mn@doi [\apj] {10.1086/159701},
  \href {https://ui.adsabs.harvard.edu/abs/1982ApJ...254...22M} {254, 22}

\bibitem[\protect\citeauthoryear{{Maraschi} \& {Tavecchio}}{{Maraschi} \&
  {Tavecchio}}{2003}]{Maraschi03}
{Maraschi} L.,  {Tavecchio} F.,  2003, \mn@doi [\apj] {10.1086/342118}, \href
  {https://ui.adsabs.harvard.edu/abs/2003ApJ...593..667M} {593, 667}

\bibitem[\protect\citeauthoryear{Maraschi et~al.,}{Maraschi
  et~al.}{2008}]{Maraschi08}
Maraschi L.,  et~al., 2008, \mn@doi [The Astrophysical Journal Letters]
  {10.1086/312370}, 526, L81

\bibitem[\protect\citeauthoryear{{Marscher} \& {Gear}}{{Marscher} \&
  {Gear}}{1985}]{Marscher85}
{Marscher} A.~P.,  {Gear} W.~K.,  1985, \mn@doi [\apj] {10.1086/163592}, \href
  {https://ui.adsabs.harvard.edu/abs/1985ApJ...298..114M} {298, 114}

\bibitem[\protect\citeauthoryear{{Mastichiadis} \& {Kirk}}{{Mastichiadis} \&
  {Kirk}}{1995}]{MK95}
{Mastichiadis} A.,  {Kirk} J.~G.,  1995, \aap, \href
  {https://ui.adsabs.harvard.edu/abs/1995A&A...295..613M} {295, 613}

\bibitem[\protect\citeauthoryear{{Mastichiadis} \& {Kirk}}{{Mastichiadis} \&
  {Kirk}}{1997}]{MK97}
{Mastichiadis} A.,  {Kirk} J.~G.,  1997, \aap, \href
  {https://ui.adsabs.harvard.edu/abs/1997A&A...320...19M} {320, 19}

\bibitem[\protect\citeauthoryear{{Mastichiadis} \& {Moraitis}}{{Mastichiadis}
  \& {Moraitis}}{2008}]{MM08}
{Mastichiadis} A.,  {Moraitis} K.,  2008, \mn@doi [\aap]
  {10.1051/0004-6361:200810505}, \href
  {https://ui.adsabs.harvard.edu/abs/2008A&A...491L..37M} {491, L37}

\bibitem[\protect\citeauthoryear{{Mastichiadis}, {Petropoulou}  \&
  {Dimitrakoudis}}{{Mastichiadis} et~al.}{2013}]{Mast2013}
{Mastichiadis} A.,  {Petropoulou} M.,   {Dimitrakoudis} S.,  2013, \mn@doi
  [\mnras] {10.1093/mnras/stt1210}, \href
  {https://ui.adsabs.harvard.edu/abs/2013MNRAS.434.2684M} {434, 2684}

\bibitem[\protect\citeauthoryear{{Mattox} et~al.,}{{Mattox}
  et~al.}{1996}]{1996ApJ...461..396M}
{Mattox} J.~R.,  et~al., 1996, \mn@doi [\apj] {10.1086/177068}, \href
  {https://ui.adsabs.harvard.edu/abs/1996ApJ...461..396M} {461, 396}

\bibitem[\protect\citeauthoryear{{Meyer}, {Scargle}  \& {Blandford}}{{Meyer}
  et~al.}{2019}]{Meyer19}
{Meyer} M.,  {Scargle} J.~D.,   {Blandford} R.~D.,  2019, \mn@doi [\apj]
  {10.3847/1538-4357/ab1651}, \href
  {https://ui.adsabs.harvard.edu/abs/2019ApJ...877...39M} {877, 39}

\bibitem[\protect\citeauthoryear{{Narayan} \& {Piran}}{{Narayan} \&
  {Piran}}{2012}]{Narayan12}
{Narayan} R.,  {Piran} T.,  2012, \mn@doi [\mnras]
  {10.1111/j.1365-2966.2011.20069.x}, \href
  {https://ui.adsabs.harvard.edu/abs/2012MNRAS.420..604N} {420, 604}

\bibitem[\protect\citeauthoryear{{Nolan} et~al.,}{{Nolan} et~al.}{2012}]{2FGL}
{Nolan} P.~L.,  et~al., 2012, \mn@doi [\apjs] {10.1088/0067-0049/199/2/31},
  \href {https://ui.adsabs.harvard.edu/abs/2012ApJS..199...31N} {199, 31}

\bibitem[\protect\citeauthoryear{{Paltani} \& {T{\"u}rler}}{{Paltani} \&
  {T{\"u}rler}}{2005}]{Paltani05}
{Paltani} S.,  {T{\"u}rler} M.,  2005, \mn@doi [\aap]
  {10.1051/0004-6361:20041206}, \href
  {https://ui.adsabs.harvard.edu/abs/2005A&A...435..811P} {435, 811}

\bibitem[\protect\citeauthoryear{{Paltani}, {Courvoisier}, {Blecha}  \&
  {Bratschi}}{{Paltani} et~al.}{1997}]{Paltani97}
{Paltani} S.,  {Courvoisier} T.~J.~L.,  {Blecha} A.,   {Bratschi} P.,  1997,
  \aap, \href {https://ui.adsabs.harvard.edu/abs/1997A&A...327..539P} {327,
  539}

\bibitem[\protect\citeauthoryear{{Peterson} et~al.,}{{Peterson}
  et~al.}{2004}]{Peterson04}
{Peterson} B.~M.,  et~al., 2004, \mn@doi [\apj] {10.1086/423269}, \href
  {https://ui.adsabs.harvard.edu/abs/2004ApJ...613..682P} {613, 682}

\bibitem[\protect\citeauthoryear{{Petropoulou}}{{Petropoulou}}{2014}]{Petro14}
{Petropoulou} M.,  2014, \mn@doi [\aap] {10.1051/0004-6361/201424603}, \href
  {https://ui.adsabs.harvard.edu/abs/2014A&A...571A..83P} {571, A83}

\bibitem[\protect\citeauthoryear{{Petropoulou} \& {Mastichiadis}}{{Petropoulou}
  \& {Mastichiadis}}{2011}]{Petropoulou11}
{Petropoulou} M.,  {Mastichiadis} A.,  2011, \mn@doi [\aap]
  {10.1051/0004-6361/201116763}, \href
  {https://ui.adsabs.harvard.edu/abs/2011A&A...532A..11P} {532, A11}

\bibitem[\protect\citeauthoryear{{Petropoulou}, {Piran}  \&
  {Mastichiadis}}{{Petropoulou} et~al.}{2015}]{PPM15}
{Petropoulou} M.,  {Piran} T.,   {Mastichiadis} A.,  2015, \mn@doi [\mnras]
  {10.1093/mnras/stv1523}, \href
  {https://ui.adsabs.harvard.edu/abs/2015MNRAS.452.3226P} {452, 3226}

\bibitem[\protect\citeauthoryear{{Petropoulou}, {Nalewajko}, {Hayashida}  \&
  {Mastichiadis}}{{Petropoulou} et~al.}{2017}]{Petropoulou2017}
{Petropoulou} M.,  {Nalewajko} K.,  {Hayashida} M.,   {Mastichiadis} A.,  2017,
  \mn@doi [\mnras] {10.1093/mnrasl/slw252}, \href
  {https://ui.adsabs.harvard.edu/abs/2017MNRAS.467L..16P} {467, L16}

\bibitem[\protect\citeauthoryear{{Pian} et~al.,}{{Pian} et~al.}{1998}]{Pian98}
{Pian} E.,  et~al., 1998, \mn@doi [\apjl] {10.1086/311083}, \href
  {https://ui.adsabs.harvard.edu/abs/1998ApJ...492L..17P} {492, L17}

\bibitem[\protect\citeauthoryear{{Rajput}, {Stalin}  \&
  {Sahayanathan}}{{Rajput} et~al.}{2020}]{Rajput20}
{Rajput} B.,  {Stalin} C.~S.,   {Sahayanathan} S.,  2020, \mn@doi [\mnras]
  {10.1093/mnras/staa2708}, \href
  {https://ui.adsabs.harvard.edu/abs/2020MNRAS.498.5128R} {498, 5128}

\bibitem[\protect\citeauthoryear{{Rani} et~al.,}{{Rani}
  et~al.}{2013}]{Rani2013}
{Rani} B.,  et~al., 2013, \mn@doi [\aap] {10.1051/0004-6361/201321058}, \href
  {https://ui.adsabs.harvard.edu/abs/2013A&A...552A..11R} {552, A11}

\bibitem[\protect\citeauthoryear{{Rees}, {Netzer}  \& {Ferland}}{{Rees}
  et~al.}{1989}]{Rees89}
{Rees} M.~J.,  {Netzer} H.,   {Ferland} G.~J.,  1989, \mn@doi [\apj]
  {10.1086/168155}, \href
  {https://ui.adsabs.harvard.edu/abs/1989ApJ...347..640R} {347, 640}

\bibitem[\protect\citeauthoryear{{Safna}, {Stalin}, {Rakshit}  \&
  {Mathew}}{{Safna} et~al.}{2020}]{Safna2020}
{Safna} P.~Z.,  {Stalin} C.~S.,  {Rakshit} S.,   {Mathew} B.,  2020, \mn@doi
  [\mnras] {10.1093/mnras/staa2622}, \href
  {https://ui.adsabs.harvard.edu/abs/2020MNRAS.498.3578S} {498, 3578}

\bibitem[\protect\citeauthoryear{{Sandrinelli}, {Covino}  \&
  {Treves}}{{Sandrinelli} et~al.}{2014}]{Sandrinelli}
{Sandrinelli} A.,  {Covino} S.,   {Treves} A.,  2014, \mn@doi [\aap]
  {10.1051/0004-6361/201321558}, \href
  {https://ui.adsabs.harvard.edu/abs/2014A&A...562A..79S} {562, A79}

\bibitem[\protect\citeauthoryear{{Sarkar}, {Gupta}, {Chitnis}  \&
  {Wiita}}{{Sarkar} et~al.}{2021}]{Sarkar2021}
{Sarkar} A.,  {Gupta} A.~C.,  {Chitnis} V.~R.,   {Wiita} P.~J.,  2021, \mn@doi
  [\mnras] {10.1093/mnras/staa3211}, \href
  {https://ui.adsabs.harvard.edu/abs/2021MNRAS.501...50S} {501, 50}

\bibitem[\protect\citeauthoryear{{Schleicher} et~al.,}{{Schleicher}
  et~al.}{2019}]{glx1}
{Schleicher} B.,  et~al., 2019, \mn@doi [Galaxies] {10.3390/galaxies7020062},
  \href {https://ui.adsabs.harvard.edu/abs/2019Galax...7...62S} {7, 62}

\bibitem[\protect\citeauthoryear{{Schmidt}}{{Schmidt}}{1963}]{Schmidt3c}
{Schmidt} M.,  1963, \mn@doi [\nat] {10.1038/1971040a0}, \href
  {https://ui.adsabs.harvard.edu/abs/1963Natur.197.1040S} {197, 1040}

\bibitem[\protect\citeauthoryear{{Sikora}, {Begelman}  \& {Rees}}{{Sikora}
  et~al.}{1994}]{Sikora94}
{Sikora} M.,  {Begelman} M.~C.,   {Rees} M.~J.,  1994, \mn@doi [\apj]
  {10.1086/173633}, \href
  {https://ui.adsabs.harvard.edu/abs/1994ApJ...421..153S} {421, 153}

\bibitem[\protect\citeauthoryear{{Sikora}, {B{\l}a{\.z}ejowski}, {Begelman}  \&
  {Moderski}}{{Sikora} et~al.}{2001}]{Sikora01}
{Sikora} M.,  {B{\l}a{\.z}ejowski} M.,  {Begelman} M.~C.,   {Moderski} R.,
  2001, \mn@doi [\apj] {10.1086/321329}, \href
  {https://ui.adsabs.harvard.edu/abs/2001ApJ...554....1S} {554, 1}

\bibitem[\protect\citeauthoryear{{Sikora}, {Stawarz}, {Moderski}, {Nalewajko}
  \& {Madejski}}{{Sikora} et~al.}{2009}]{Sikora09}
{Sikora} M.,  {Stawarz} {\L}.,  {Moderski} R.,  {Nalewajko} K.,   {Madejski}
  G.~M.,  2009, \mn@doi [\apj] {10.1088/0004-637X/704/1/38}, \href
  {https://ui.adsabs.harvard.edu/abs/2009ApJ...704...38S} {704, 38}

\bibitem[\protect\citeauthoryear{{Sobrino Figaredo} et~al.,}{{Sobrino Figaredo}
  et~al.}{2020}]{Sobrino20}
{Sobrino Figaredo} C.,  et~al., 2020, \mn@doi [\aj] {10.3847/1538-3881/ab89b1},
  \href {https://ui.adsabs.harvard.edu/abs/2020AJ....159..259S} {159, 259}

\bibitem[\protect\citeauthoryear{{Soldi} et~al.,}{{Soldi} et~al.}{2008}]{Soldi}
{Soldi} S.,  et~al., 2008, \mn@doi [\aap] {10.1051/0004-6361:200809947}, \href
  {https://ui.adsabs.harvard.edu/abs/2008A&A...486..411S} {486, 411}

\bibitem[\protect\citeauthoryear{{Stawarz} \& {Kirk}}{{Stawarz} \&
  {Kirk}}{2007}]{StaKirk}
{Stawarz} {\L}.,  {Kirk} J.~G.,  2007, \mn@doi [\apjl] {10.1086/518417}, \href
  {https://ui.adsabs.harvard.edu/abs/2007ApJ...661L..17S} {661, L17}

\bibitem[\protect\citeauthoryear{Summerlin \& Baring}{Summerlin \&
  Baring}{2011}]{Summerlin11}
Summerlin E.~J.,  Baring M.~G.,  2011, \mn@doi [The Astrophysical Journal]
  {10.1088/0004-637x/745/1/63}, 745, 63

\bibitem[\protect\citeauthoryear{{Takahashi} et~al.,}{{Takahashi}
  et~al.}{1996}]{Takahashi96}
{Takahashi} T.,  et~al., 1996, \mn@doi [\apjl] {10.1086/310302}, \href
  {https://ui.adsabs.harvard.edu/abs/1996ApJ...470L..89T} {470, L89}

\bibitem[\protect\citeauthoryear{{Tavecchio} \& {Ghisellini}}{{Tavecchio} \&
  {Ghisellini}}{2008}]{tavecchio08}
{Tavecchio} F.,  {Ghisellini} G.,  2008, \mn@doi [\mnras]
  {10.1111/j.1365-2966.2008.13072.x}, \href
  {https://ui.adsabs.harvard.edu/abs/2008MNRAS.386..945T} {386, 945}

\bibitem[\protect\citeauthoryear{{Tavecchio} et~al.,}{{Tavecchio}
  et~al.}{2001}]{Tavecchio01}
{Tavecchio} F.,  et~al., 2001, \mn@doi [\apj] {10.1086/321394}, \href
  {https://ui.adsabs.harvard.edu/abs/2001ApJ...554..725T} {554, 725}

\bibitem[\protect\citeauthoryear{{Tchekhovskoy}, {Narayan}  \&
  {McKinney}}{{Tchekhovskoy} et~al.}{2011}]{Tchek11}
{Tchekhovskoy} A.,  {Narayan} R.,   {McKinney} J.~C.,  2011, \mn@doi [\mnras]
  {10.1111/j.1745-3933.2011.01147.x}, \href
  {https://ui.adsabs.harvard.edu/abs/2011MNRAS.418L..79T} {418, L79}

\bibitem[\protect\citeauthoryear{{Thiersen}, {Zacharias}  \&
  {B{\"o}ttcher}}{{Thiersen} et~al.}{2019}]{glx2}
{Thiersen} H.,  {Zacharias} M.,   {B{\"o}ttcher} M.,  2019, \mn@doi [Galaxies]
  {10.3390/galaxies7010035}, \href
  {https://ui.adsabs.harvard.edu/abs/2019Galax...7...35T} {7, 35}

\bibitem[\protect\citeauthoryear{{Ulrich}}{{Ulrich}}{1981}]{Ulrich81}
{Ulrich} M.~H.,  1981, \mn@doi [\ssr] {10.1007/BF00353556}, \href
  {https://ui.adsabs.harvard.edu/abs/1981SSRv...28...89U} {28, 89}

\bibitem[\protect\citeauthoryear{{Urry} \& {Padovani}}{{Urry} \&
  {Padovani}}{1995}]{Urry1995}
{Urry} C.~M.,  {Padovani} P.,  1995, \mn@doi [\pasp] {10.1086/133630}, \href
  {https://ui.adsabs.harvard.edu/abs/1995PASP..107..803U} {107, 803}

\bibitem[\protect\citeauthoryear{{Vaughan}, {Edelson}, {Warwick}  \&
  {Uttley}}{{Vaughan} et~al.}{2003}]{Vaughan03}
{Vaughan} S.,  {Edelson} R.,  {Warwick} R.~S.,   {Uttley} P.,  2003, \mn@doi
  [\mnras] {10.1046/j.1365-2966.2003.07042.x}, \href
  {https://ui.adsabs.harvard.edu/abs/2003MNRAS.345.1271V} {345, 1271}

\bibitem[\protect\citeauthoryear{{Wehrle} et~al.,}{{Wehrle}
  et~al.}{2012}]{Wehrle2012}
{Wehrle} A.~E.,  et~al., 2012, \mn@doi [\apj] {10.1088/0004-637X/758/2/72},
  \href {https://ui.adsabs.harvard.edu/abs/2012ApJ...758...72W} {758, 72}

\bibitem[\protect\citeauthoryear{{Wehrle} et~al.,}{{Wehrle}
  et~al.}{2016}]{Wehrle16}
{Wehrle} A.~E.,  et~al., 2016, \mn@doi [\apj] {10.3847/0004-637X/816/2/53},
  \href {https://ui.adsabs.harvard.edu/abs/2016ApJ...816...53W} {816, 53}

\bibitem[\protect\citeauthoryear{{Williamson} et~al.,}{{Williamson}
  et~al.}{2016}]{Williamson2016}
{Williamson} K.,  et~al., 2016, \mn@doi [Galaxies] {10.3390/galaxies4040064},
  \href {https://ui.adsabs.harvard.edu/abs/2016Galax...4...64W} {4, 64}

\bibitem[\protect\citeauthoryear{Wise \& Bristow-Johnson}{Wise \&
  Bristow-Johnson}{1999}]{Wise99}
Wise D.,  Bristow-Johnson R.,  1999, in Performance of Low-Order Polynomial
  Interpolators in the Presence of Oversampled Input.

\bibitem[\protect\citeauthoryear{{Yoshida}, {Bailyn}, {Cruz}, {Urry}, {Coppi},
  {Vasilopoulos}, {Petropoulou}  \& {Meyer}}{{Yoshida}
  et~al.}{2020}]{Yoshida2020}
{Yoshida} K.,  {Bailyn} C.,  {Cruz} B.,  {Urry} C.,  {Coppi} P.,
  {Vasilopoulos} G.,  {Petropoulou} M.,   {Meyer} M.,  2020, in American
  Astronomical Society Meeting Abstracts \#235. p. 405.08

\bibitem[\protect\citeauthoryear{{Yoshida}, {Bailyn}, {Urry}  \& {et
  al.}}{{Yoshida} et~al.}{2021}]{Yoshida2021}
{Yoshida} K.,  {Bailyn} C.,  {Urry} C.~M.,   {et al.} 2021, in preparation

\bibitem[\protect\citeauthoryear{{Zacharias} \& {Schlickeiser}}{{Zacharias} \&
  {Schlickeiser}}{2010}]{Zacharias10}
{Zacharias} M.,  {Schlickeiser} R.,  2010, \mn@doi [\aap]
  {10.1051/0004-6361/201015284}, \href
  {https://ui.adsabs.harvard.edu/abs/2010A&A...524A..31Z} {524, A31}

\bibitem[\protect\citeauthoryear{{Zhang} et~al.,}{{Zhang}
  et~al.}{2019}]{Zhang19}
{Zhang} Z.-X.,  et~al., 2019, \mn@doi [\apj] {10.3847/1538-4357/ab1099}, \href
  {https://ui.adsabs.harvard.edu/abs/2019ApJ...876...49Z} {876, 49}

\bibitem[\protect\citeauthoryear{Zhang, Gupta, Gaur, Wiita, An, Lu, Fan  \&
  Xu}{Zhang et~al.}{2021}]{Zhang_2021}
Zhang Z.,  Gupta A.~C.,  Gaur H.,  Wiita P.~J.,  An T.,  Lu Y.,  Fan S.,   Xu
  H.,  2021, \mn@doi [The Astrophysical Journal] {10.3847/1538-4357/abdd38},
  909, 103

\bibitem[\protect\citeauthoryear{{de Hoffmann} \& {Teller}}{{de Hoffmann} \&
  {Teller}}{1950}]{deHoffmann50}
{de Hoffmann} F.,  {Teller} E.,  1950, \mn@doi [Physical Review]
  {10.1103/PhysRev.80.692}, \href
  {https://ui.adsabs.harvard.edu/abs/1950PhRv...80..692D} {80, 692}

\makeatother
\end{thebibliography}



\appendix
\section{Analytical considerations for the estimation of $\sigma_\gamma$}\label{sec:app}
In this section we provide simple analytical arguments for the dependence of the $\gamma$-ray flux on the time-varying model parameters, namely $l_{\rm e}, B, l_{\rm ext}$, and $\delta$. We follow \cite{PPM15} and present their main findings that are relevant to our work. 

\cite{PPM15} computed the steady-state synchrotron and SSC photon compactnesses\footnote{We define the photon compactness of emission component $j$ with energy density $u_{\rm j}$ as $l_{\rm j} \equiv \sigma_{\rm T} R u_{\rm j}/m_{\rm e}c^2$.} ($l_{\rm syn}$ and $l_{\rm ssc}$ respectively) as a function of the injection electron compactness $l_{\rm e}$ for mono-energetic electrons with Lorentz factor $\gamma_{\rm e}$, approximating  the single-particle emissivities with $\delta$-functions. These authors examined the steady-state solutions in the following characteristic regimes:
\begin{itemize}
    \item {\it Slow-cooling regime.} Electron cooling due to synchrotron and ICS processes is negligible and the steady-state electron distribution is determined by the balance between the escape and injection processes. The synchrotron and Compton compactnesses are written as:
    \begin{eqnarray}
    l_{\rm syn} & = & 4  \gamma_{\rm e} l_{\rm e} l_{\rm b} \\
    l_{\rm ssc} & = & \left(4  \gamma_{\rm e} l_{\rm e}\right)^2 l_{\rm b} \\
    l_{\rm ecs} & = & 4 \gamma_{\rm e}  l_{\rm e}  l_{\rm ext} 
    \end{eqnarray}
    where we introduced the compactness of the ECS component ($l_{\rm ecs}$) and $l_{\rm b}=\sigma_{\rm T} B^2 / 8 \pi m_{\rm e} c^2$. 
    \item {\it Fast-cooling regime.} Electron cooling due to synchrotron or ICS processes is important and the steady-state electron distribution is determined by the balance between the energy loss and injection processes. In this case, the compactness of the various emission components is given by 
    \begin{eqnarray}
    l_{\rm syn} & = & \frac{1}{2} l_{\rm b} f_{\rm ext}\left(-1+\sqrt{1+\frac{12 l_{\rm e}}{l_{\rm b} f^2_{\rm ext}}}\right) \\
    l_{\rm ssc} & = & \left( \frac{l_{\rm syn}}{l_{\rm b}} \right)^2 l_{\rm b} \\
    l_{\rm ecs} & = & \frac{1}{2} l_{\rm ext} f_{\rm ext}\left(-1+\sqrt{1+\frac{12 l_{\rm e}}{l_{\rm b} f^2_{\rm ext}}}\right). 
    \end{eqnarray}
    where $f_{\rm ext} \equiv 1+ l_{\rm ext}/ l_{\rm b}$.
\end{itemize}
Noting that the $\gamma$-ray compactness, $l_{\gamma} \propto L_{\gamma}/R$, is equal to $l_{\rm ssc (ecs)}$ in the SSC (ECS) scenario, while $l_{\rm syn}$ is a proxy of the OIR non-thermal photon compactness $l_{\rm oir}$, we derive the following scalings for the two emission scenarios:
\begin{itemize}
    \item {\it SSC scenario.} Here, $l_{\rm ext}=0$ and $f_{\rm ext}=1$.
    \begin{eqnarray}
    l_{\rm oir} \propto 
    \begin{cases}
    l_{\rm e} B^2, & {\rm slow \, cooling} \\
    l_{\rm e}, & {\rm fast \, cooling \, \& \, } 12 l_{\rm e} \ll l_{\rm b}\\
    l^{1/2}_{\rm e} B, & {\rm fast \, cooling \, \& \, } 12 l_{\rm e} \gg l_{\rm b}.
    \end{cases}    
    \label{eq:ssc-opt}
    \end{eqnarray}
    \begin{eqnarray}
    l_{\gamma} \propto 
    \begin{cases}
    l_{\rm e}^2 B^2, & {\rm slow \, cooling} \\
    l_{\rm e}^2 B^{-2}, & {\rm fast \, cooling \, \& \, } 12 l_{\rm e} \ll l_{\rm b}\\
    l_{\rm e}, & {\rm fast \, cooling \, \& \, } 12 l_{\rm e} \gg l_{\rm b}.
    \end{cases}    
    \label{eq:ssc}
    \end{eqnarray}
    For the $l_{\rm e}$-varying simulations $\sigma_\gamma=2$ or 1 depending on the cooling regime and the dominant cooling process (synchrotron or SSC, respectively). Similarly, for the $B$-varying simulations $\sigma_\gamma=2, -2$ or 0.
    \item {\it ECS scenario.} Here, $f_{\rm ext} >1$.
    \begin{eqnarray}
    l_{\rm oir} \propto 
    \begin{cases}
    l_{\rm e} B^2, & {\rm slow \, cooling} \\
    l_{\rm e}, & {\rm fast \, cooling \, \& \, } 12 l_{\rm e} \ll l_{\rm b} f^2_{\rm ext}  \, \& \, l_{\rm ext} \ll l_{\rm b}\\
    l_{\rm e} l_{\rm ext}^{-1}B^2, & {\rm fast \, cooling \, \& \, } 12 l_{\rm e} \ll l_{\rm b} f^2_{\rm ext}  \, \& \, l_{\rm ext} \gg l_{\rm b}\\
    l_{\rm e}^{1/2}B, & {\rm fast \, cooling \, \& \, } 12 l_{\rm e} \gg l_{\rm b} f^2_{\rm ext}.   
    \end{cases}    
    \label{eq:ecs-opt}
    \end{eqnarray}
    \begin{eqnarray}
    l_{\gamma} \propto 
    \begin{cases}
    l_{\rm e} l_{\rm ext}, & {\rm slow \, cooling} \\
    l_{\rm e} l_{\rm ext} B^{-2}, & {\rm fast \, cooling \, \& \, } 12 l_{\rm e} \ll l_{\rm b} f^2_{\rm ext}  \, \& \, l_{\rm ext} \ll l_{\rm b}\\
    l_{\rm e}, & {\rm fast \, cooling \, \& \, } 12 l_{\rm e} \ll l_{\rm b} f^2_{\rm ext}  \, \& \, l_{\rm ext} \gg l_{\rm b}\\
    l_{\rm e}^{1/2}l_{\rm ext}B^{-1}, & {\rm fast \, cooling \, \& \, } 12 l_{\rm e} \gg l_{\rm b} f^2_{\rm ext}.   
    \end{cases}    
    \label{eq:ecs}
    \end{eqnarray}
For the $l_{\rm ext}$-varying simulations we find $\sigma_\gamma=1$ except for the extreme case where electrons are fast cooling due to ECS and the $\gamma$-ray flux becomes independent of $l_{\rm ext}$ (and $B$). For the $l_{\rm e}$-varying simulations and most parameter combinations, we find $\sigma_\gamma=1$. Finally, for the $B$-varying simulations we expect  $\sigma_\gamma=-2, -1$ or 0 depending on the cooling regime and the dominant cooling mechanism. 
\end{itemize}    

For the bolometric $\gamma$-ray luminosity emitted by a relativistically moving blob, $L_{\gamma, \rm obs}\propto \delta^4 l_{\gamma}$ \citep{Ghisellini13}. When considering the $\gamma$-ray luminosity in a specific energy range, as in our numerical simulations, spectral shifts due to variations in $\delta$ may become important and deviations from $\sigma_\gamma=4$ are expected.

To summarize, this analysis does not take into account spectral changes due to particle cooling or extended particle distributions that may become important when studying fluxes in a narrow energy range. Nonetheless, the analytical scalings presented above serve as a roadmap for the selection of the $\sigma_\gamma$ value in our numerical calculations (see Table~\ref{tab:sigmag}). From this analysis, it becomes also clear that a choice of a constant value $\sigma_\gamma$ value is a simplification, because the radiating particles may change cooling regimes throughout the simulations.


\bsp	
\label{lastpage}
\end{document}